\documentclass[aps,preprint,groupedaddress, nofootinbib]{revtex4}

\usepackage{amsmath,amsfonts,amssymb,amsthm, bbm}

\usepackage{enumerate}  
\usepackage[pdftex]{graphicx} 
\usepackage{rotating}  

\newcommand\herm{\mathbb H(\mathcal H)}
\newcommand\unit{\mathbb U(\mathcal H)}
\newcommand\den{\mathbb D(\mathcal H)}
\newcommand\eff{\mathbb E(\mathcal H)}
\newcommand\proj{\mathbb P(\mathcal H)}
\newcommand\povm{\mathrm{POVM}(\mathcal H)}
\newcommand\pvm{\mathrm{PVM}(\mathcal H)}

\newcommand\Tr{\mathrm{Tr}}

\newcommand\spec{\mathrm{spec}}
\newcommand\Hil{\mathcal H}
\newcommand{\SU}{\mathrm{SU}(2)}
\newcommand{\id}{\mathbbm 1}

\newcommand\ip[2]{\langle#1, #2\rangle}
\newcommand\abs[1]{|#1|}
\newcommand\norm[1]{\|#1\|}
\newcommand{\field}[1]{\mathbb {#1}}
\newcommand\w[1]{e^{\frac{2\pi i}{d}#1}}

\newcommand\pure{\mathrm{pure}}

\def\tD{\mathtt D}
\def\tF{\mathtt F}
\def\fF{\mathfrak F}

\newtheorem{definition}{Definition}
\newtheorem{theorem}{Theorem}
\newtheorem{lemma}{Lemma}
\newtheorem{proposition}{Proposition}

\usepackage{hyperref} 

\begin{document}

\title{Framed Hilbert space: hanging the quasi-probability pictures of quantum theory}
\author{Christopher Ferrie}
\affiliation{
Institute for Quantum Computing,
University of Waterloo,
Waterloo, Ontario, Canada, N2L 3G1}
\affiliation{
Department of Applied Mathematics,
University of Waterloo,
Waterloo, Ontario, Canada, N2L 3G1}
\author{Joseph Emerson}
\affiliation{
Institute for Quantum Computing,
University of Waterloo,
Waterloo, Ontario, Canada, N2L 3G1}
\affiliation{
Department of Applied Mathematics,
University of Waterloo,
Waterloo, Ontario, Canada, N2L 3G1}

\begin{abstract}
Building on earlier work, we further develop a formalism based on
the mathematical theory of frames that defines a set of possible
phase-space or quasi-probability representations of
finite-dimensional quantum systems.  We prove that an alternate approach to defining a set of quasi-probability representations, based on a more natural generalization of a classical representation, is equivalent to our earlier approach based on frames, and therefore is also subject to our no-go theorem for a non-negative representation. Furthermore, we clarify the relationship between the
contextuality of quantum theory and the necessity of negativity in
quasi-probability representations and discuss their relevance as
criteria for non-classicality.  We also provide a comprehensive
overview of known quasi-probability representations and their
expression within the frame formalism.
\end{abstract}

\date{\today}

\maketitle


\tableofcontents

\section{Intoduction\label{sec:intro}}

One approach to establish a common ground between classical and
quantum theory is phase space. Phase space is a natural concept in
classical theory since it is equivalent to the state space. The idea
of formulating quantum theory in phase space dates back to the early
days of quantum theory when the so-called Wigner function was
introduced \cite{Wig32}.  The Wigner function is a
\emph{quasi-probability} distribution on a classical phase space\footnote{Note that the Wigner function is not the only such
function.  A review of the Wigner function and related
representations appears in \cite{Lee94}.}. The term
quasi-probability refers to the fact that the function is not a true
probability density as it takes on negative values for some quantum
states. The Wigner function formalism can be lifted into a fully
autonomous phase space theory which reproduces all the predictions
of the standard quantum theory of infinite dimensional systems
\cite{Bak58}.  In other words, this phase space formulation of
quantum theory is equivalent to the usual abstract formalism of
quantum theory in the same sense that Heisenberg's matrix mechanics
and Schrodinger's wave mechanics are equivalent to the abstract
formalism.

However, the Wigner function representation is non-unique, and,
moreover, Wigner's approach is not applicable to describing quantum
systems with a finite set of distinguishable states. In recent years
various analogs of the Wigner function for finite dimensional
quantum systems have been proposed. An important subclass of such
quasi-probability representations are those defined on a
\emph{discrete phase space}. We will use the term
\emph{quasi-probability representation} to refer to the broader
class of representations where the (ontic) state space is neither
required to be discrete nor required to carry any classical phase
space structure\footnote{The term \emph{quasi-probability} is used
to allow for the appearance of negative values in the states and/or
measurements defined under the representation.}. Such
representations have provided insight into fundamental properties of
finite-dimensional quantum systems. For example, the representation
proposed by Wootters identifies sets of mutually unbiased bases
\cite{Woo87,GHW04}. Inspired by the discovery that quantum resources
lead to algorithms that dramatically outperform their classical
counterparts, there has also been growing interest in the
application of discrete phase space formalism to analyze the
quantum-classical contrast for finite-dimensional systems.  Examples
include analyses of quantum teleportation \cite{Paz02}, the effect
of decoherence on quantum walks \cite{LP03}, quantum Fourier
transform and Grover's algorithm \cite{MPS02}, conditions for
exponential quantum computational speedup \cite{Gal05,CGG06}, and
quantum expanders \cite{GJ07}.

As noted above, a central concept in studies of the
quantum-classical contrast in the quasi-probability formalisms of
quantum theory is the appearance of \emph{negativity}. A
non-negative quasi-probability function is a true probability
distribution, prompting some authors to suggest that the presence of
negativity in this function is a defining signature of
non-classicality. Unfortunately the application of any one of these
quasi-probability representations in the context of determining
criteria for the non-classicality of a given quantum task is limited
in significance by the non-uniqueness of that particular
representation. Ideally one would like to determine whether the task
can be expressed as a classical process in \emph{any}
quasi-probability representation. Indeed the sheer variety of
proposed quasi-probability representations prompts the question of
whether there is some shared underlying mathematical structure that
might provide a means for identifying the full family of such
representations.  Moreover, from an operational view, states alone
are an incomplete description of an experimental arrangement. Hence,
it is important to elucidate the ways in which a quasi-probability
representation \emph{of states alone} can be lifted into an
autonomous quasi-probability representation of \emph{both the states
and measurements} defining any set of experimental configurations.

In an earlier paper \cite{FE08}, we introduced the concept of
\emph{frame representation} which mathematically unifies the known
quasi-probability representations of quantum states.  Once the frame
is identified for a particular quasi-probability representation, we
showed how any dual frame leads to an autonomous representation of
the operational quantum formalism (i.e., including both states
\emph{and} measurements). Then we proved that such representation
must possess negative values in either the states or measurements
(or both). However, because this approach explicitly made use of
dual frames to construct the autonomous formulation of quantum
mechanics, it remained possible that negativity could be avoided
through some other
representation that is not explicitly defined via a selection of a dual frame. In this paper we show that this is not
possible. Specifically, we consider an alternative and more natural approach to the definition of the class of autonomous quasi-probability representations for finite-dimensional quantum mechanics and show
that any such representation is equivalent to a frame
representation and therefore can not avoid negativity.

The outline of the paper is as follows. In Section \ref{sec:review
quasi prob functions}, we give a comprehensive overview of the known
quasi-probability functions representing quantum states.  In Section
\ref{chap:Frame_states}, we review the results of \cite{FE08}
showing how the mathematical theory of frames provides a formalism
which underlies all known quasi-probability representations of
finite dimensional quantum \emph{states}.  Furthermore, we explicitly
construct the frames which give rise to several important
quasi-probability representations.  In Section
\ref{sec:quasi-probabilityrepsof} we review the two ways, presented
in \cite{FE08}, in which any frame representation of states can be
extended to include a representation of measurements, and hence
lifted to a fully autonomous formulation of finite dimensional
quantum mechanics. In Section \ref{sec:necessity of negativity} we propose a more natural approach to the definition of quasi-probability representations and show that this approach is equivalent to the approach obtained under frame representations and therefore does not admit a non-negative representation.  In Section \ref{chap:contextuality}, we look at the
equivalence of non-negativity in quasi-probability
representations and non-contextuality in ontological models. In Section \ref{sec:conclusion} we discuss the extension
of our results to infinite-dimensional quantum systems and consider
potential applications of our work in quantum information science.

\section{Review of quasi-probability functions\label{sec:review quasi prob functions}}
Reviewed in this section are the existing quasi-probability
representations of quantum states found in the literature. The
original quasi-probability representation put forth by Wigner and
later recognized as an equivalent formulation of the full quantum
theory by Moyal \cite{Moy49} and others \cite{Bak58} is reviewed
first.  This \emph{phase space} picture is only valid for infinite
dimensional Hilbert spaces but it will be reviewed here because it
has motivated all known generalizations for finite dimensions.
Sections \ref{sec:Wigner}-\ref{sec:Fields} are devoted to reviewing
a representative sample of the known quasi-probability
representations of finite dimensional quantum states.

In Section \ref{sec:qp_summary} a summary of these and a few other quasi-probability representations is presented in a more concise manner.

\subsection{Wigner phase space representation\label{sec:Wigner}}

The position operator $Q$ and momentum operator $P$ are
the central objects in the abstract formalism of infinite
dimensional quantum theory.  The operators satisfy the canonical
commutation relations
\begin{equation*}
[  Q,   P]=i.\label{CCR}
\end{equation*}
Since $  Q$ and $  P$ do not commute, the choice of the quantization
map $g(q,p)\mapsto g(  Q,  P)$, for some function $g$, is not unique.  This is the so-called ``ordering
problem''.  A class of solutions to this problem is the association
$e^{i\xi q+i\eta p}\mapsto e^{i\xi   Q+i\eta   P}f(\xi,\eta)$ for some
arbitrary function $f$ (See Table 1 of \cite{Lee94} for a review of
the traditional choices for $f$).

Consider the classical particle phase space $\field R^2$ and the continuous set of operators
\begin{equation}
  F(q,p):=\frac{1}{(2\pi)^2}\int_{\field R^2} d\xi d\eta
\;e^{i\xi(q-  Q)+i\eta(p-  P)}f(\xi,\eta).\label{Wigner_phasepoints}
\end{equation}
When $f(\xi,\eta)=1$, the distribution
\begin{align}
\mu_\rho^{\textrm{Wigner}}(q,p)&:=\Tr( \rho  F(q,p))\label{def_Wigner}\\
&=\frac{1}{(2\pi)^2}\int_{\field R^2} d\xi d\eta \;\Tr\left[  \rho
e^{i\xi(  Q-q)+i\eta(  P-p)}\right]
\end{align}
is the celebrated Wigner function \cite{Wig32}.  The Wigner function is both positive and negative in general.  However, it otherwise behaves as a classical probability density on the classical phase space.  For these reasons, the Wigner function and others like it came to be called \emph{quasi-probability} functions.

The Wigner function is the unique representation satisfying the properties \cite{BB87}
\begin{enumerate}[(a)]
\item[Wig(1)] For all $ \rho$, $\mu_{\rho}^{\textrm{Wigner}}(q,p)$ is real.
\item[Wig(2)] For all $ \rho_1$ and $ \rho_2$, $$\Tr( \rho_1 \rho_2)=2\pi \int_{\field R^2} d\xi d\eta\; \mu_{\rho_1}^{\textrm{Wigner}}(\xi,\eta)\mu_{\rho_2}^{\textrm{Wigner}}(\xi,\eta).$$
\item[Wig(3)] For all $ \rho$, integrating $\mu_{\rho}^{\textrm{Wigner}}$ along the line $a  q+\ b p=c$ in phase space yields the probability that a measurement of the observable $a   Q+ b  P$ has the result $c$.
\end{enumerate}
Notice from Equation \eqref{def_Wigner} that Wigner function is obtained from the set of operators in Equation \eqref{Wigner_phasepoints} (for $f=1$) via the trace.  Thus the properties Wig(1)-(3) can be transformed into properties on a set of operators $  F(q,p)$ which uniquely specify the set in Equation \eqref{Wigner_phasepoints} for $f=1$.  These properties are
\begin{enumerate}[(a)]
\item[Wig(4)] $  F(q,p)$ is Hermitian.
\item[Wig(5)] $2\pi\Tr(F(q,p)F(q',p'))=\delta(q-q')\delta(p-p')$.
\item[Wig(6)] Let $  P_c$ be the projector onto the eigenstate of $a  Q+b  P$ with eigenvalue $c$.  Then, $$\int_{\field R^2} dq dp\; F(q,p)\delta(aq+bp-c)=  P_c.$$
\end{enumerate}

The Wigner functions has many properties and applications \cite{HOSW84} which are not of concern here.  However, it is important to note that the wide variety of fruitful applications of the Wigner function is responsible for the interest in its generalization.  The properties Wig(1)-(6) were presented here as most authors have aimed at a finite dimensional analogy of the Wigner function defined such that it satisfies properties analogous to Wig(1)-(6) for discrete phase spaces.  The remainder of the section is devoted to generalizing the definition of the Wigner function to finite dimensional quantum systems.

\subsection{Wootters discrete phase space representation\label{sec:Wootters}}

In \cite{Woo87}, Wootters is interested in obtaining a discrete
analog of the Wigner function.  Associated with each Hilbert space $\Hil$ of finite dimension $d$ is a
\emph{discrete phase space}.  First assume $d$ is
prime.  The \emph{prime phase space}, $\Phi_d$, is
a $d\times d$ array of points $\alpha=(q,p)\in \mathbb
Z_d\times\mathbb Z_d$.

A \emph{line}, $\lambda$, is the set of
$d$ points satisfying the linear equation $aq+bp=c$, where all
arithmetic is modulo $d$.  Two lines are \emph{parallel} if their
linear equations differ in the value of $c$.  The prime phase space $\Phi_d$ contains $d+1$ sets of $d$ parallel
lines called \emph{striations}.

Assume the the Hilbert space $\Hil$ has composite dimension
$d=d_1d_2\cdots d_k$.  The discrete phase space of the entire $d$
dimensional system is the Cartesian product of two-dimensional prime
phase spaces of the subsystems. The phase space is thus a $(d_1\times
d_1) \times (d_2\times d_2)\times\cdots(\times d_k\times d_k)$ array.
Such as construction is formalized as follows: the
\emph{discrete phase space} is the multi-dimensional array
$\Phi_d=\Phi_{d_1}\times\Phi_{d_2}\times\cdots\times\Phi_{d_k}$,
where each $\Phi_{d_i}$ is a prime phase space.  A \emph{point} is
the $k$-tuple $\alpha=(\alpha_1,\alpha_2, \ldots, \alpha_k)$ of
points $\alpha_i=(q_i,p_i)$ in the prime phase spaces.  A
\emph{line} is the $k$-tuple
$\lambda=(\lambda_1,\lambda_2,\ldots,\lambda_k)$ of lines in the
prime phase spaces.  That is, a line is the set of $d$ points
satisfying the equation
$$(a_1q_1+b_1p_1,a_2q_2+b_2p_2,\ldots,a_kq_k+b_kp_k)=(c_1,c_2,\ldots,c_k),$$
which is symbolically written $aq+bp=c$.  Two lines are
\emph{parallel} if their equations differ in the value
$c$.  As was the case for the prime phase spaces, parallel lines can be partitioned into sets, again called striations; the
discrete phase space $\Phi_d$ contains $(d_1+1)(d_2+1)\cdots(d_k+1)$ sets of $d$
parallel lines.

The construction of the discrete phase space is now been complete.
To introduce Hilbert space into the discrete phase space formalism,
Wootters chooses the following special basis for the space of Hermitian operators.  The set of operators $\{  A_\alpha:\alpha\in\Phi_d\}$ acting on an $d$ dimensional
Hilbert space are called \emph{phase point operators} if the operators satisfy
\begin{enumerate}[(a)]
\item[Woo(4)] For each point $\alpha$, $A_\alpha$ is Hermitian.
\item[Woo(5)] For any two points $\alpha$ and $\beta$, $\Tr(  A_\alpha  A_\beta)=d\delta_{\alpha\beta}$.
\item[Woo(6)] For each line $\lambda$ in a given striation, the operators
$  P_\lambda=\frac{1}{d}\displaystyle\sum_{\alpha\in\lambda}
A_\alpha$ form a projective valued measurement (PVM): a set of $d$
orthogonal projectors which sum to identity.
\end{enumerate}
Notice that these properties of the phase point operators Woo(4)-(6) are discrete analogs of the properties Wig(4)-(6) of the function $F$ defining the original Wigner function.  This definition suggests that the lines in the discrete phase
space should be labeled with states of the Hilbert space.  Since each striation is
associated with a PVM, each of the $d$ lines in a striation is
labeled with an orthogonal state.  For each $\Phi_d$, there is a unique set of phase point operators up
to unitary equivalence.

Although the sets of phase point operators are unitarily equivalent,
the induced labeling of the lines associated to the chosen set of
phase point operators are not equivalent.  This is clear from the
fact that unitarily equivalent PVMs do not project onto the same
basis.

The choice of phase point operators in
\cite{Woo87} will be adopted.
For $d$ prime, the phase point operators are
\begin{equation}
  A_\alpha=\frac{1}{d}\sum_{j,m=0}^{d-1}
\omega^{pj-qm+\frac{jm}{2}}  X^j  Z^m,\label{Wooters_phasepoint_prime}
\end{equation}
where $\omega$ is a $d$'th root of unity and $  X$ and $  Z$ are the generalized Pauli operators (see Appendix \ref{sec:mathematical notations}). For
composite $d$, the phase point operator in $\Phi_d$ associated with
the point $\alpha=(\alpha_1,\alpha_2,\ldots,\alpha_k)$ is given by
\begin{equation}
  A_\alpha=  A_{\alpha_1}\otimes
A_{\alpha_2}\otimes\cdots\otimes
A_{\alpha_k},\label{Wootters_phasepoint_composite}
\end{equation}
where each $  A_{\alpha_i}$ is the phase point operator of
the point $\alpha_i$ in $\Phi_{d_i}$.

The $d^2$ phase point operators are linearly independent and form a
basis for the space of Hermitian operators acting on an $d$ dimensional
Hilbert space.  Thus, any density operator $  \rho$ can be
decomposed as
\[
  \rho=\sum_{q,p} \mu_\rho^{\textrm{Wootters}}(q,p)  A(q,p),
\]
where the real coefficients are explicitly given by
\begin{equation}
\mu_\rho^{\textrm{Wootters}}(q,p)=\frac{1}{d}\Tr(  \rho  A(q,p)).\label{rho2DWF}
\end{equation}
This discrete phase space function
is the Wootters \emph{discrete Wigner function}.  This discrete quasi-probability function satisfies the following properties which are the discrete analogies of the properties Wig(1)-(3) the original continuous Wigner function satisfies.
\begin{enumerate}[(a)]
\item[Woo(1)] For all $ \rho$, $\mu_{\rho}^{\textrm{Wootters}}(q,p)$ is real.
\item[Woo(2)] For all $ \rho_1$ and $ \rho_2$, $$\Tr( \rho_1 \rho_2)= d \sum_{q,p} \mu_{\rho_1}^{\textrm{Wootters}}(q,p)\mu_{\rho_2}^{\textrm{Wootters}}(q,p).$$
\item[Woo(3)] For all $ \rho$, summing $\mu_{\rho}^{\textrm{Wootters}}$ along the line $\lambda$ in phase space yields the probability that a measurement of the PVM associated with the striation which contains $\lambda$ has the outcome associated with $\lambda$.
\end{enumerate}

\subsection{Odd dimensional discrete Wigner functions\label{sec:Odd}}

In \cite{CCSS87}, Cohendet \emph{et al} define a discrete analogue of the Wigner function which is valid for integer spin.  That is, $\dim(\Hil)=d$ is assumed to be odd.  Whereas Wootters builds up a discrete phase space before defining a Wigner function, the authors of \cite{CCSS87} implicitly define a discrete phase space through the definition of their Wigner function.

Consider the operators
\begin{equation*}
  W_{mn}\phi_k=\omega^{2n(k-m)}\phi_{k-2m},
\end{equation*}
with $m,n\in\field Z_d$ and ${\phi_k}$ are the eigenvectors of $Z$ (see Appendix \ref{sec:mathematical notations}).  Then, the \emph{discrete Wigner function} of a density operator $ \rho$ is
\begin{equation}\label{DWF_odd}
\mu_\rho^{\textrm{odd}}(q,p)=\frac{1}{d}\Tr( \rho  W_{qp}  P),
\end{equation}
where $  P$ is the parity operator (see Appendix \ref{sec:mathematical notations}).

The authors call the operators $ \triangle_{qp}=  W_{qp}  P$ \emph{Fano operators} and note that they satisfy
\begin{align*}
& \triangle_{qp}^\dag= \triangle_{qp},\\
& \triangle_{qp}^2= \id,\\
&\Tr( \triangle_{qp} \triangle_{q'p'})=d\delta_{qq'}\delta_{pp'},\\
&  W_{xk}^\dag \triangle_{qp}  W_{xk}= \triangle_{q-2x\;p-2k}.
\end{align*}
The Fano operators play a role similar to Wootters' phase point operators; they form a complete basis of the space of Hermitian operators.  The phase space implicitly defined through the definition of the discrete Wigner function \eqref{DWF_odd} is $\field Z_d\times\field Z_d$.  When $d$ is an odd prime, this phase space is equivalent to Wootters discrete phase space.  In this case the Fano operators are $ \triangle_{qp}=A_{(-q,p)}$.  This can seen by writing the Wootters phase point operators as
\begin{equation*}
A_{(q,p)}=\frac{1}{d}  X^{2q}  Z^{2p}  P \omega^{2qp}.
\end{equation*}

\subsection{Even dimensional discrete Wigner functions\label{sec:Even}}

In \cite{Leo95}, Leonhardt defines discrete analogues of the Wigner function for both odd and even dimensional Hilbert spaces.  In a later paper \cite{Leo96}, Leonhardt discusses the need for separate definitions for the odd and even dimension cases.   Naively applying his definition, or that of Cohendet \emph{et al}, of the discrete Wigner function for odd dimensions to even dimensions yields unsatisfactory results.  The reason for this is the discrete Wigner function carries redundant information for even dimensions which is insufficient to specify the state uniquely.  The solution is to enlarge the phase space until the information in the phase space function becomes sufficient to specify the state uniquely.

Suppose $\dim(\Hil)=d$ is odd.  Leonhardt defines the discrete Wigner function as
\begin{equation*}
\mu_\rho^{\textrm{Leonhardt}}(q,p)=\frac{1}{d}\Tr( \rho  X^{2q}  Z^{2p}   P \omega^{2qp}).
\end{equation*}
Leonhardt's definition of an odd dimensional discrete Wigner function is unitarily equivalent to the Cohendet \emph{et al} definition.  That is, $\mu_\rho^{\textrm{Leonhardt}}(q,p)=\mu_\rho^{\textrm{odd}}(-q,p)$.  To define a discrete Wigner function for even dimensions, Leonhardt takes half-integer values of $q$ and $p$.  This amounts to enlarging the phase space to $\field Z_{2d}\times\field Z_{2d}$.  Thus the \emph{even dimensional} discrete Wigner function is
\begin{equation*}
\mu_\rho^{\textrm{even}}(q,p)= \frac{1}{2d}\Tr( \rho  X^{q}  Z^{p}   P \omega^{\frac{qp}{2}}),
\end{equation*}
where the operators
\[
\triangle^{\textrm{even}}_{qp}=\frac{1}{2d}  X^{q}  Z^{p}   P \omega^{\frac{qp}{2}}
\]
could be called the even dimensional Fano or phase point operators.  Of course, these operators do not satisfy all the criteria which the Fano operators (in the case of Cohendet \emph{et al}) or the phase point operators (in the case of Wootters) satisfy; they are not orthogonal.  Moreover, they are not even linearly independent which can easily be inferred since there are $4d^2$ of them and a set of linearly independent operators contains a maximum of $d^2$ operators.

\subsection{Wigner functions on the sphere\label{sec:Spehere}}

In \cite{HW00}, Heiss and Weigert are concerned with a set of postulates put forth by Stratonovich \cite{Str57}.   The aim of Stratonovich was to find a Wigner function type mapping, analogous to that of a infinite dimensional system on $\field R^2$, of a $d$ dimensional system on the sphere $\field S^2$.  The first postulate is linearity and is always satisfied if the Wigner functions on the sphere satisfy
\begin{equation}
\mu_\rho^{\textrm{sphere}}(\mathbf{n})=\Tr( \rho  \triangle(\mathbf{n})),\label{DWF_sphere}
\end{equation}
where $\mathbf{n}$ is a point on $\field S^2$.  The remaining postulates on this quasi-probability mapping are
\begin{align}
&\mu_\rho^\textrm{sphere}(\mathbf{n})^\ast=\mu_\rho^\textrm{sphere}(\mathbf{n}),\nonumber\\
&\frac{d}{4\pi}\int_{\field S^2} d\mathbf{n} \;\mu_\rho^\textrm{sphere}(\mathbf{n})=1,\nonumber\\
&\frac{d}{4\pi}\int_{\field S^2} d\mathbf{n} \;\mu_{\rho_1}^\textrm{sphere}(\mathbf{n})\mu_{\rho_2}^\textrm{sphere}(\mathbf{n})=\Tr( \rho_1 \rho_2),\nonumber\\
&\mu_{(g\cdot\rho)}^\textrm{sphere}(\mathbf{n})=\mu_\rho^\textrm{sphere}(\mathbf{n})^g,\;g\in\SU,\nonumber
\end{align}
where $g\cdot\rho$ is the image of $  U_g \rho  U_g^\dag$ and $  U:\SU\to \unit$ is an irreducible unitary representation of the group $\SU$.  These postulates are analogous to Wig(1)-(3) for the Wigner function modulo the second normalization condition (which could have be included in the Wigner function properties).

The continuous set of operators $ \triangle(\mathbf{n})$ is called a \emph{kernel} and plays the role of the phase point and Fano operators of the previous sections.  Requiring that Equation \eqref{DWF_sphere} hold changes the postulates to new conditions on the kernel
\begin{align}
& \triangle(\mathbf{n})^\dag= \triangle(\mathbf{n}),\label{kernel_1}\\
&\frac{d}{4\pi}\int_{\field S^2} d\mathbf{n} \; \triangle(\mathbf{n})= \id,\label{kernel_2}\\
&\frac{d}{4\pi}\int_{\field S^2} d\mathbf{n} \;\Tr( \triangle(\mathbf{n}) \triangle(\mathbf{m})) \triangle(\mathbf{n})= \triangle(\mathbf{m}),\label{kernel_3}\\
& \triangle(g\cdot\mathbf{n})=  U_g \triangle(\mathbf{n})  U_g^\dag,\;g\in\SU.\label{kernel_4}
\end{align}
These postulates are the spherical analogies of properties Wig(4)-(6) (again, modulo the normalization condition).
Heiss and Weigert provide a derivation of $2^{2s}$, where  $s=\frac{d-1}{2}$ is the \emph{spin}, unique kernels satisfying these postulates.  They are
\begin{equation}\label{cont_kernel}
 \triangle(\mathbf{n})=\sum_{m=-s}^{s}\sum_{l=0}^{2s}\epsilon_l \frac{2l+1}{2s+1}C^{s\;l\;s}_{m\;0\;m} \phi_m(\mathbf{n})\phi_m^\ast(\mathbf{n}),
\end{equation}
where $C$ denotes the so-called \emph{Clebsch-Gordon coefficients}; $\phi_m(\mathbf{n})$ are the eigenvectors of the operator $\mathbf{  S}\cdot\mathbf{n}$, where $\mathbf{  S}=(  X,  Y,  Z)$; and $\epsilon_l=\pm1$, for $l=1\ldots 2s$ and $\epsilon_0=1$.

Heiss and Weigert relax the postulates Equations \eqref{kernel_1}-\eqref{kernel_4} on the kernel $ \triangle(\mathbf{n})$ to allow for a pair of kernels $ \triangle^\mathbf{n}$ and $ \triangle_\mathbf{m}$.  The pair individually satisfy Equation \eqref{kernel_1}, while one of them satisfies Equation \eqref{kernel_2} and the other Equation \eqref{kernel_4}.  Together, the pair must satisfy the generalization of Equation \eqref{kernel_3}
\begin{equation}
\frac{d}{4\pi}\int_{\field S^2} d\mathbf{n}\Tr( \triangle^\mathbf{n} \triangle_\mathbf{m}) \triangle^\mathbf{n}= \triangle_\mathbf{m}.\label{kernel_dual}
\end{equation}
A pair of kernels, together satisfying Equation \eqref{kernel_dual}, is given by
\begin{align*}
 \triangle_\mathbf{n}&=\sum_{m=-s}^{s}\sum_{l=0}^{2s}\gamma_l \frac{2l+1}{2s+1}C^{s\;l\;s}_{m\;0\;m} \phi_m(\mathbf{n})\phi_m^\ast(\mathbf{n}),\\
 \triangle^\mathbf{n}&=\sum_{m=-s}^{s}\sum_{l=0}^{2s}\gamma_l^{-1} \frac{2l+1}{2s+1}C^{s\;l\;s}_{m\;0\;m} \phi_m(\mathbf{n})\phi_m^\ast(\mathbf{n}),
\end{align*}
where $\gamma_l=\pm1$ for $l=1\ldots 2s$ and $\gamma_0=1$.  The original postulates are satisfied when $\gamma_l=\gamma_l^{-1}\equiv\epsilon_l$.

The major contribution of \cite{HW00} is the derivation of a \emph{discrete} kernel $ \triangle_\nu:= \triangle_{\mathbf{n}_\nu}$, for $\nu=1\ldots d^2$ which satisfies the discretized postulates
\begin{align}
& \triangle_\nu^\dag= \triangle_\nu,\label{discrete_kernel_1}\\
&\frac{1}{d}\sum_{\nu=1}^{d^2}  \triangle^\nu= \id,\label{discrete_kernel_2}\\
&\frac{1}{d}\sum_{\nu=1}^{d^2} \Tr( \triangle_\nu \triangle^\mu) \triangle_\nu= \triangle^\mu,\label{discrete_kernel_3}\\
& \triangle_{g\cdot \nu}=  U_g \triangle_\nu  U_g^\dag,\;g\in\SU.\label{discrete_kernel_4}
\end{align}
The subset of points $\mathbf{n}_\nu$ is called a \emph{constellation}.  The linearity postulate is not explicitly stated since it is always satisfied under the assumption
\begin{equation}
 \rho\to\mu_\rho^{\textrm{constellation}}(\nu)=\Tr( \rho  \triangle_\nu).\label{DWF_constellation}
\end{equation}
Equation \eqref{discrete_kernel_3} is called a \emph{duality} condition.  That is, it is only satisfied if $ \triangle_\nu$ and $ \triangle^\mu$ are \emph{dual bases} for $\herm$.  In particular,
\begin{equation*}
\frac{1}{d}\Tr( \triangle_\nu \triangle^\mu)=\delta_{\nu\mu}.
\end{equation*}
Although the explicit construction of a pair of discrete kernels satisfying Equations \eqref{discrete_kernel_1}-\eqref{discrete_kernel_4} might be computationally hard, their existence is a trivial exercise in linear algebra.  Indeed, so long as $ \triangle_\nu$ is a basis for $\herm$, its dual, $ \triangle^\mu$, is uniquely determined by
\begin{equation*}
 \triangle^\mu=\sum_{\nu=1}^{d^2} \mathtt{G}^{-1}_{\nu\mu} \triangle_\nu,
\end{equation*}
where the Gram matrix $\mathtt G$ is given by
\begin{equation*}
\mathtt G_{\nu\mu}=\Tr( \triangle_\nu \triangle_\mu).
\end{equation*}
The authors of \cite{HW00} note that almost any constellation leads to a discrete kernel $ \triangle_\nu$ forming a basis for $\herm$.  The term \emph{almost any} here means that a randomly selected discrete kernel will form, with probability 1, a basis for $\herm$.

\subsection{Finite fields discrete phase space representation\label{sec:Fields}}
Recall that when $\dim(\Hil)=d$ is prime, Wootters defines the discrete phase space as a $d\times d$ lattice indexed by the group $\field Z_d$.  In \cite{Woo04}, Wootters generalizes his original construction of a discrete phase space to allow the $d\times d$ lattice to be indexed by a finite field $\field F_d$ which exists when $d=p^n$ is an integer power of a prime number.  This approach is discussed at length in the paper \cite{GHW04} authored by Gibbons, Hoffman and Wootters (GHW).

Similar to his earlier approach, Wootters defines the \emph{phase space}, $\Phi_d$, as
a $d\times d$ array of points $\alpha=(q,p)\in \mathbb
F_d\times\mathbb F_d$.  A \emph{line}, $\lambda$, is the set of
$d$ points satisfying the linear equation $aq+bp=c$, where all
arithmetic is done in $\field F_d$.  Two lines are \emph{parallel} if their
linear equations differ in the value of $c$.

The mathematical structure of $\mathbb{F}_d$ is appealing because
lines defined as above have the following useful properties: (i)
given any two points, exactly one line contains both points, (ii)
given a point $\alpha$ and a line $\lambda$ not containing $\alpha$,
there is exactly one line parallel to $\lambda$ that contains
$\alpha$, and (iii) two nonparallel lines intersect at exactly one
point. Note that these are usual properties of lines in Euclidean
space.  As before, the $d^2$ points of the phase space $\Phi_d$ can be partitioned into
$d+1$ sets of $d$ parallel lines called \emph{striations}.  The
line containing the point $(q,p)$ and the origin $(0,0)$ is called a
\emph{ray} and consists of the points $(sq,sp)$, where $s$ is a
parameter taking values in $\mathbb{F}_d$.  We choose each ray,
specified by the equation $aq+bp=0$, to be the representative of the
striation it belongs to.

A translation in phase space, $\mathcal T_{\alpha_0}$, adds a constant
vector, $\alpha_0=(q_0,p_0)$, to every phase space point:
$\mathcal T_{\alpha_0}\alpha=\alpha+\alpha_0$.  Each line, $\lambda$, in a
striation is invariant under a translation by any point contained in
its ray, parameterized by the points $(sq,sp)$. That is,
\begin{equation}
\tau_{(sq,sp)}\lambda=\lambda.\label{striaeinv}
\end{equation}

The discrete Wigner function is
\begin{equation*}
\mu_\rho^{\textrm{field}}(q,p)=\frac{1}{d}\Tr( \rho A_{(q,p)}),
\end{equation*}
where now the Hermitian \emph{phase point operators} satisfy the following properties for a projector valued function $  Q$, called a \emph{quantum net}, to be defined later.
\begin{enumerate}[(a)]
\item[GHW(4)] For each point $\alpha$, $  A$ is Hermitian.
\item[GHW(5)] For any two points $\alpha$ and $\beta$, $\Tr(  A_\alpha  A_\beta)=d\delta_{\alpha\beta}$.
\item[GHW(6)] For any line $\lambda$, $\displaystyle\sum_{\alpha\in\lambda}A_\alpha=d   Q(\lambda)$.
\end{enumerate}
The projector valued function $  Q$ assigns quantum states to lines in phase space.  This mapping is required to satisfy the special property of \emph{translational covariance}, which is defined after a short, but necessary, mathematical digression.  Notice first that properties GHW(4) and GHW(5) are identical to Woo(4) and Woo(4).  Also note that if GHW(6) is to be analogous to Woo(6), the property of translation covariance must be such that the set $\{  Q(\lambda)\}$ when $\lambda$ ranges over a striation forms a PVM.

The set of elements $E=\{e_0,...,e_{n-1}\}\subset \field F_d$ is called a
\emph{field basis} for $\field{F}_d$ if any element, $x$, in
$\field{F}_d$ can be written
\begin{equation}
x=\sum_{i=0}^{n-1}x_ie_i,\label{fieldbasis}
\end{equation}
where each $x_i$ is an element of the prime field $\mathbb{Z}_p$.
The \emph{field trace}\footnote{Note that we will distinguish the
field trace, $\textrm{tr}(\cdot)$, from the usual trace of a Hilbert
space operator, $\textrm{Tr}(\cdot)$, by the case of the first
letter.} of any field element is given by
\begin{equation}
\textrm{tr}(x)=\sum_{i=0}^{n-1}x^{p^i}.\label{fieldtrace}
\end{equation}
There exists a unique field basis, $\tilde E =\{\tilde e_0,...\tilde
e_{n-1}\}$, such that $\textrm{tr}(\tilde e_i e_j)=\delta_{ij}$.  We
call $\tilde E$ the \emph{dual} of $E$.

The construction presented in \cite{GHW04} is physically significant for a
system of $n$ objects (called \emph{particles}) having a $p$
dimensional Hilbert space. A translation operator, $  T_\alpha$
associated with a point in phase space $\alpha=(q,p)$ must act
independently on each particle in order to preserve the tensor
product structure of the composite system's Hilbert space.  We
expand each component of the point $\alpha$ into its field basis
decomposition as in Equation \eqref{fieldbasis}
\begin{equation}
q=\sum_{i=0}^{n-1}q_ie_i\label{q}
\end{equation}
and
\begin{equation}
p=\sum_{i=0}^{n-1}p_if\tilde e_i,\label{p}
\end{equation}
with $f$ any element of $\mathbb{F}_d$.  Note that the basis we
choose for $p$ is a multiple of the dual of that chosen for $q$.
Now, the translation operator associated with the point $(q,p)$ is
\begin{equation}
  T_{(q,p)}=\bigotimes_{i=0}^{n-1}  X^{q_i}  Z^{p_i},\label{transopt}
\end{equation}
Since $X$ and $Z$ are unitary, $  T_\alpha$ is unitary.

We assign with each line in phase space a pure quantum state.  The
quantum net $  Q$ is defined such that for
each line, $\lambda$, $  Q(\lambda)$ is the operator which projects
onto the pure state associated with $\lambda$.  As a consequence of the
choice of basis for $p$ in Equation \eqref{p}, the state assigned to the line
$\tau_\alpha\lambda$ is obtained through
\begin{equation}
  Q(\tau_\alpha\lambda)=  T_\alpha
  Q(\lambda)  T_\alpha^\dag.\label{transcov}
\end{equation}
This is the condition of translational covariance and it implies that each striation is associated with an orthonormal
basis of the Hilbert space.  To see this, recall the property in Equation
\eqref{striaeinv}. From Equation \eqref{transcov}, this implies that, for
each $s\in\mathbb{F}_d$, $  T_{(sq,sp)}$ must commute with
$  Q(\lambda)$, where the line $\lambda$ is any line in the
striation defined by the ray consisting of the points $(sq,sp)$.
That is, the states associated to the lines of the striation must be
common eigenstates of the unitary translation operator
$  T_{(sq,sp)}$, for each $s\in \mathbb{F}_d$. Thus, the states are
orthogonal and form a basis for the Hilbert space.  That is, their projectors form a PVM which makes GHW(6) identical to Woo(6) when $d$ is prime.

In \cite{GHW04}, the author's note that, although the association
between states and vertical and horizontal lines is fixed, the
quantum net is not unique.  In fact, there are $d^{d-1}$ quantum
nets which satisfy Equation \eqref{striaeinv}.  When $d$ is prime, one of these quantum nets corresponds exactly to the original discrete Wigner function defined by Wootters in Section \ref{sec:Wootters}.

\subsection{Summary of existing quasi-probability representations of quantum states\label{sec:qp_summary}}
The phase space functions reviewed in Sections
\ref{sec:Wootters}-\ref{sec:Fields} form only a subset of the
literature on finite dimensional phase space functions; there are
indeed several others (for a recent review see \cite{Vou04}).  More
generally, there exist \emph{quasi-probability representations}
given by real-valued representations that do not necessarily reflect
any preconceived classical phase space structure.  For example in
\cite{Har01} Hardy shows that five axioms are sufficient to imply a
special quasi-probabilistic representation which is equivalent to an
operational form of quantum theory. In \cite{Hav03} Havel also
proposes an kind of analog of the Wigner function called the ``real
density matrix''.

An overview of all of the quasi-probability representations for
finite dimensional quantum systems reviewed above (as well as a
couple more) is presented in Table \ref{table:qp}.  The table
identifies the ontic space structure and the mathematical field
which indexes it (if applicable).  The second to last column
indicates whether or not the representation contains redundant
information. The last column reveals the scope of quantum theory the
paper aims to cover (notice that typically only states are
considered).

\begin{sidewaystable}[ht]
\caption{Finite quasi-probability representations}
\begin{centering}
\begin{tabular}{c c c c c c c}
\hline\hline
Author(s) & Year & Valid dimensions & Phase space & Index field & Redundancy & Quantum theory scope\\ [0.5ex] 
\hline 
Stratonovich \cite{Str57} & 1957 & any & sphere & polar coordinates & continuous & states\\
Wootters \cite{Woo87} & 1987 & prime* & $d\times d$ lattice & $\mathbb Z_d$ & no & standard\\
Cohendet \emph{et al} \cite{CCSS87} & 1987 & odd & $d\times d$ lattice & $\mathbb Z_d$ & no & states\\
Leonhardt \cite{Leo95} & 1995 & even** & $2d\times 2d$ lattice &   $\mathbb Z_{2d}$ & four-fold & states\\
Heiss and Weigert \cite{HW00} & 2000 & any & sphere*** &   arbitrary & no & states\\
Hardy \cite{Har01} & 2001 & any & none & n/a & no & operational\\
Havel \cite{Hav03} & 2003 & any & none & n/a & no & states\\
Gibbons \emph{et al} \cite{GHW04} & 2004 & power of prime & $d\times d$ lattice & $\mathbb F_d$ & no & states\\
Ruzzi \emph{et al} \cite{RMG05} & 2005 & odd & $d\times d$ lattice & $\mathbb Z_d$ & no & states\\
Chaturvedi \emph{et al} \cite{CEMMMS06} & 2006 & any & $d\times d$ lattice & $\mathbb Z_d$ & no & states\\
Gross \cite{Gro06} & 2006 & odd & $d\times d$ lattice & $\mathbb Z_d$ & no & states\\
[1ex] 
\hline 
\end{tabular}
\end{centering}\\
Notes: *Wootters' original discrete Wigner function \cite{Woo87} is usually understood to be valid for prime dimension but, as discussed in Section \ref{sec:Wootters}, is easily extended to any dimension by combining prime dimensional phase spaces.  **Leonhardt \cite{Leo95} also defines a discrete Wigner function valid for odd dimensional which is equivalent to the other odd dimensional cases \cite{CCSS87}.  ***The phase space of Heiss and Weigert is any subset of $d$ points on the sphere which can indexed arbitrarily.
\label{table:qp} 
\end{sidewaystable}

\section{Frame representations of quantum states\label{chap:Frame_states}}
In the previous section we reviewed some of the quasi-probability
representations of quantum states found in the literature.  In
Section \ref{sec:qp_unification}, we unify these known
quasi-probability representations of quantum states into one concise
mathematical definition.

Section \ref{sec:intro_frames} contains an introduction to the
mathematical theory of frames.  In Section \ref{sec:Equivalence} it
is shown that frame theory is both necessary and sufficient to
define any quasi-probability representation of quantum states.  The
frames for each of the quasi-probability representations reviewed in
Section \ref{sec:review quasi prob functions} are given in Section
\ref{sec:Examples}.

\subsection{Unification of existing quasi-probability representations of quantum states\label{sec:qp_unification}}

Each of the quasi-probability functions discussed in Section \ref{sec:review quasi prob functions} are \emph{linear} representations of the density operator.  These representations are also \emph{invertible} as the density operator can be obtained from any quasi-probability function.  Each quasi-probability function is a also member of the function space $L^2(\Lambda)$, where $\Lambda$ represents a classical state space (e.g. the phase space, where applicable).  These three properties will constitute the following minimal definition of a quasi-probability representation of \emph{quantum states alone} \cite{FE08}:
\begin{definition}\label{def:quasiprobstates}
A \emph{quasi-probability representation of quantum states} is any map
$\herm\to L^2(\Lambda)$ that is linear and invertible.
\end{definition}
This definition was chosen since all the known quasi-probability representations of states satisfy it.  One might object that the restrictions imposed on this map are too strong.  Indeed there is no mention of linearity, invertibility, or $L^2$ spaces in any notion of classical probability.  However, a classical probabilistic description describes an entire experimental arrangement.  In Section \ref{sec:equivaffineqp} we show that these ad hoc requirements follow from more natural assumptions on a representation on a complete experimental description.  That is, Definition \ref{def:quasiprobstates} is satisfied on the part of this more natural definition which represents quantum states.

Given Definition \ref{def:quasiprobstates}, any phase space function is
then a particular type of quasi-probability representation.
\begin{definition}\label{def:phase_space_states}
If there exists a symmetry group on $\Lambda$, $G$, carrying a
unitary representation $  U:G\to\unit$ and  a
quasi-probability representation satisfying the covariance property
$  U_g\rho  U_g^\dag\mapsto \{\mu_\rho(g(\lambda))\}_{\lambda\in\Lambda}$ for
all $\rho\in\den$ and $g\in G$, then $  \rho\mapsto \mu_\rho(\lambda)$ is a \emph{phase space
representation of quantum states}.
\end{definition}
All phase space functions in the literature
correspond to quasi-probability representations that satisfy this additional covariance condition.

Table \ref{table:qp} shows that the range of validity in the Hilbert space dimension of these functions are often disjoint.  Moreover, the construction of the phase spaces use varying mathematical structures: integers, finite fields and points on a sphere.  Coupled with the fact that at least two known representations required redundancy, it may seem at first that Definition \ref{def:quasiprobstates} is as far as one can can go in unifying the quasi-probability functions.  However, a much stronger result is possible.  In \cite{FE08} it was shown that the mathematical theory of \emph{frames} is both sufficient \emph{and necessary} to describe any representation of quantum states satisfying Definition \ref{def:quasiprobstates}.

\subsection{The mathematical theory of frames\label{sec:intro_frames}}

A \emph{frame} can be thought of as a
generalization of an orthonormal basis \cite{Chr03}.  However, the particular Hilbert space under consideration here is not $\Hil$.  Considered here is a generalization of a basis for $\herm$, which is the set of Hermitian operators on an complex Hilbert space of dimension $d$.  With the trace inner product (or Hilbert-Schmidt inner product) $\ip{  A}{  B}:=\Tr(  A  B)$, $\herm$ forms a Hilbert space itself of dimension $d^2$.  Let $\Lambda$ be some measure space\footnote{For brevity, the $\sigma$-algebra and measure are left implied.}.
\begin{definition}\label{def:frame}
A frame for $\herm$ is a set of operators $\mathcal F:=\{ F(\lambda)\}\subset\herm$ which satisfies
\begin{equation}\label{def_discrete_frame}
a\norm{A}^2\leq\int_\Lambda d\lambda\;\abs{\ip{ F(\lambda)}{A}}^2\leq b\norm{A}^2,
\end{equation}
for all $A\in\herm$ and some constants $a,b>0$.
\end{definition}
This definition generalizes a defining condition for an orthogonal basis $\{ B_k\}_{k=1}^{d^2}$
\begin{equation}\label{def_basis}
\sum_{k=1}^{d^2}\abs{\ip{ B_k}{A}}^2 = \norm{A}^2,
\end{equation}
for all $A\in\herm$.
\begin{definition}\label{def:dual}
A frame $\mathcal D:=\{ D(\lambda)\}$ which satisfies
\begin{equation}\label{def_dual}
A=\int_\Lambda d\lambda\; \ip{ F(\lambda)}{A} D(\lambda),
\end{equation}
for all $A\in\herm$, is a \emph{dual frame} (to $\mathcal F$).
\end{definition}
The \emph{frame operator} associated with the frame $\mathcal F$ is defined as
\begin{equation*}\label{def_frameop}
 S(A):=\int_\Lambda d\lambda\;  \ip{ F(\lambda)}{A}F(\lambda).
\end{equation*}
If the frame operator satisfies $ S=a\tilde \id$, the frame is
called \emph{tight}. The frame operator is invertible and thus every
operator has a representation
\begin{align}\label{frame_decomposition}
A= S^{-1} SA=\int_\Lambda d\lambda\; \ip{ F(\lambda)}{A} S^{-1} F(\lambda).
\end{align}
The frame $ S^{-1}\mathcal F$ is called
the \emph{canonical dual frame}.  When $\abs {\Lambda}=d^2$, the canonical dual frame is the unique
dual, otherwise there are infinitely many choices for a
dual.

A tight frame is ideal from the perspective that its canonical dual is proportional to the frame itself.  Hence, the reconstruction is given by the convenient formula
\begin{align*}
A= S^{-1} SA=\frac{1}{a}\int_\Lambda d\lambda\; \ip{ F(\lambda)}{A} F(\lambda)
\end{align*}
which is to be compared with
\begin{align*}
A=\sum_{k=1}^{d^2}\ip{ B_k}{A} B_k
\end{align*}
which defines $\{ B_k\}_{k=1}^{d^2}$ as an orthonormal basis.

The mapping $A\mapsto\ip{ F(\lambda)}{A}$ is usually called the \emph{analysis operation} in the frame literature as it encodes the signal in terms of the frame.  Here the notion of a signal is not appropriate and a more suggestive name has been chosen and formalized in the following definition.
\begin{definition}\label{def:framerep}
A mapping $\herm\to L^2(\Lambda)$ of the form
\begin{equation}\label{def_frame_rep}
  A\mapsto \ip{ F(\lambda)}{A},
\end{equation}
where $\{ F(\lambda)\}$ is a frame, is a \emph{frame representation} of $\herm$.
\end{definition}

\subsection{Equivalence of the quasi-probability and frame representation of quantum states\label{sec:Equivalence}}
Since each frame has at least a canonical dual, a frame representation (Definition \ref{def:framerep}) can always be inverted according to the reconstruction formula in Equation \eqref{frame_decomposition}.  A frame representation is defined such that it exists in $L^2(\Lambda)$.  It is clear that a frame representation is linear by virtue of the linearity of the inner product.  Thus each frame representation is guaranteed to be a quasi-probability representation of quantum states (Definition \ref{def:quasiprobstates}).  However, it is not immediately clear that the converse is true: every quasi-probability representation of quantum states is a frame representation.  Indeed, the following lemma establishes the equivalence between frame representations and quasi-probability representations of quantum states.
\begin{lemma}\label{lemma_frame}
A mapping $\herm\to L^2(\Lambda)$ is quasi-probability representation of quantum states (Definition \ref{def:quasiprobstates})
if and only if it is a frame representation for some unique frame $\mathcal F$.
\end{lemma}
\begin{proof}
The proof of this lemma also appears in \cite{FE08}.  As noted above, it is clear that a frame representation is a quasi-probability representation. So assume the mapping $W:\herm\to L^2(\Lambda)$ is a quasi-probability representation.  Linearity and the Riesz
representation theorem imply that
$W(A)(\lambda)=\ip{F(\lambda)}{A}$ for some unique set
$\mathcal F:=\{F(\lambda)\}$ (not necessarily a frame).  Since $\herm$ is finite dimensional, the inverse $W^{-1}$ is bounded.  Thus $W$ is bounded below by the bounded inverse theorem. That is, there exists a constant $a>0$ such that
\begin{equation*}
a\norm{A}^2\leq\int_\Lambda d\lambda\abs{\ip{F(\lambda)}{A}}^2.
\end{equation*}
Since $\ip{F(\lambda)}{A}\in L^2(\Lambda)$, there exists a constant $b>0$ such that
\begin{equation*}
\int_\Lambda d\lambda\abs{\ip{F(\lambda)}{A}}^2\leq b\norm{A}^2.
\end{equation*}
Hence $\mathcal F$ is a frame.
\end{proof}

Thus there is a unique frame which defines each of the quasi-probability functions reviewed in Section \ref{sec:review quasi prob functions}.  In the cases where the representation of the density operator is not redundant, the frame is just a basis.  In the redundant cases, Leonhardt's even dimensional representation (Section \ref{sec:Even}) for example, the formalism of frame theory is necessary as a basis will not suffice.

\subsection{Examples of quasi-probability representations of quantum states\label{sec:Examples}}

The frames for each of the quasi-probability representations of quantum states reviewed in
Section \ref{sec:review quasi prob functions} are now given.

\subsubsection{Wootters discrete Wigner function\label{sec:Example_Wootters}}

Let $d$ be a prime number.  Here $\Lambda=\field Z_d\times\field Z_d$.  Consider the frame $\mathcal F^{\textrm{Wootters}}=\{F^{\textrm{Wootters}}(q,p)\}$, where
\begin{equation*}
F^{\textrm{Wootters}}(q,p)=\frac{1}{d^2}  X^{2q}  Z^{2p}  P \omega^{2qp}.\label{Woottersframe}
\end{equation*}
The quasi-probability function $\mu_\rho^{\textrm{Wootters}}$ is a frame representation given by the frame $\mathcal F^{\textrm{Wootters}}$.  The frame operator of $\mathcal F^{\textrm{Wootters}}$ is $ S=d^{-1}\tilde \id$.  The unique dual frame of $\mathcal F^{\textrm{Wootters}}$ is given
by $ S^{-1}\mathcal F^{\textrm{Wootters}}$, where here $ S^{-1}=d\tilde \id$. Comparing this result to Equation \eqref{Wooters_phasepoint_prime}, the
dual frame to $\mathcal F^{\textrm{Wootters}}$ is a set of phase point
operators.

Consider the group of translations on $\Lambda$ with unitary representation $T_{(q,p)}=X^qZ^p$.  Then,
\begin{align*}
T_{(q,p)}F^{\textrm{Wootters}}(q',p') T^\dag_{(q,p)}&=\frac{1}{d^2}X^{q}  Z^p  X^{2q'}  Z^{2p'}  P Z^{-p}X^{-q}\omega^{2q'p'}\\
&=\frac{1}{d^2}  X^{2(q+q')}  Z^{2(p+p')}  P\omega^{2(q+q')(p+p')}\\
&=F^{\textrm{Wootters}}(q+q',p+p').
\end{align*}
Thus, by definition, the Wootters representation is a phase space representation.

Recall from Section \ref{sec:Wootters} that Wootters also considered non-prime dimensions.  In that case, the phase point operators (Equation \eqref{Wootters_phasepoint_composite}) were a tensor product of phase point operators (Equation \eqref{Wooters_phasepoint_prime}) for prime dimensions.  The same is true here for the frame in composite dimensions.  When $d$ is composite with prime decomposition $d=d_1d_2\cdots d_k$.  Let $\Lambda=\Lambda_1\times\Lambda_2\times\cdots\Lambda_k$ where each $\Lambda_i=\field Z_{d_i}\times\field Z_{d_i}$.
When $d$ is composite the frame is $\mathcal F^{\textrm{Wootters}}=\{F^{\textrm{Wootters}}(q_{(i)},p_{(i)})\}$ where
\[
F^{\textrm{Wootters}}(q_{(i)},p_{(i)})=F^{\textrm{Wootters}}(q_{(1)},p_{(1)})\otimes F^{\textrm{Wootters}}(q_{(2)},p_{(2)})\otimes \cdots\otimes F^{\textrm{Wootters}}(q_{(k)},p_{(k)})
\]
and each $F^{\textrm{Wootters}}(q_{(i)},p_{(i)})$ is a frame as in Equation \eqref{Woottersframe}.

\subsubsection{Odd dimensional discrete Wigner functions\label{sec:Example_Odd}}

Let $d$ be an odd integer.  Here $\Lambda=\field Z_d\times\field Z_d$.  Consider the frame $\mathcal F^{\textrm{odd}}=\{F^{\textrm{odd}}(q,p)\}$, where
\begin{equation*}
F^{\textrm{odd}}(q,p)=\frac{1}{d^2}  X^{-2q}  Z^{2p}  P \omega^{-2qp}.\label{oddframe}
\end{equation*}
The quasi-probability function $\mu_\rho^{\textrm{odd}}$ is a frame representation given by the frame $\mathcal F^{\textrm{odd}}$.
The frame operator of $\mathcal F^{\textrm{odd}}$ is
$ S=d^{-1}\tilde \id$.  The unique dual frame of $\mathcal F^{\textrm{odd}}$ is given
by $ S^{-1}\mathcal F^{\textrm{odd}}$, where here $ S^{-1}=d\tilde \id$. The
dual frame to $\mathcal F^{\textrm{odd}}$ is what Cohendet \emph{et al.} call a set of Fano
operators.

Consider the group of translations on $\Lambda$ with unitary representation $T_{(q,p)}=X^{-q} Z^p$.  Then,
\begin{align*}
T_{(q,p)}F^{\textrm{odd}}(q',p')T^\dag_{(q,p)}=F^{\textrm{odd}}(q+q',p+p').
\end{align*}
By definition, this odd dimensional representation is a phase space representation.

\subsubsection{Even dimensional discrete Wigner functions\label{sec:Example_Even}}

Let $d$ be an even integer.  Then $\Lambda=\field Z_{2d}\times\field Z_{2d}$.  Consider the frame $\mathcal F^{\textrm{even}}=\{F^{\textrm{even}}(q,p)\}$, where
\begin{equation*}
F^{\textrm{even}}(q,p)=\frac{1}{4d^2}  X^{q}  Z^{p}   P \omega^{\frac{qp}{2}}.\label{evenframe}
\end{equation*}
The quasi-probability function $\mu_\rho^{\textrm{even}}$ is a frame representation given by the frame $\mathcal F^{\textrm{even}}$.
The frame operator of $\mathcal F^{\textrm{even}}$ is
$ S=(2d)^{-1}\tilde \id$.  However, this implies the frame is only tight; it is not a basis and the dual is not unique.  The canonical dual frame of $\mathcal F^{\textrm{even}}$ is given
by $ S^{-1}\mathcal F^{\textrm{even}}$, where here $ S^{-1}=2d\tilde \id$.  The canonical
dual frame to $\mathcal F^{\textrm{even}}$ is what was called a set of even dimensional Fano
operators.

Consider the group of translations on $\Lambda$ with unitary representation $ T_{(q,p)}=X^\frac{q}{2} Z^\frac{p}{2}$.  Then,
\begin{align*}
 T_{(q,p)}F^{\textrm{even}}(q',p') T^\dag_{(q,p)}=F^{\textrm{even}}(q+q',p+p').
\end{align*}
Thus, by definition, this odd dimensional representation is a phase space representation.

\subsubsection{Wigner functions on the sphere\label{sec:Example_Spehere}}

Let $d$ be any integer.  Here the phase space is $\Lambda=\field S^2$.  Consider the frame $\mathcal F^{\textrm{sphere}}:=\{F^{\textrm{sphere}}(\mathbf{n})\}$ given by
\begin{equation*}\label{sphereframe}
  F^{\textrm{sphere}}(\mathbf{n})=  \triangle(\mathbf{n}),
\end{equation*}
where $  \triangle(\mathbf{n})$ is the same kernel given in  Equation \eqref{cont_kernel}.
The quasi-probability function $\mu_\rho^{\textrm{sphere}}$ is a frame representation given by the frame $\mathcal F^{\textrm{sphere}}$.  From Equation \eqref{kernel_3}, it follows that the frame operator of $\mathcal F^{\textrm{sphere}}$ is $ S={4\pi}d^{-1}\tilde \id$.  Thus the frame is tight.  Equation \eqref{kernel_4} is the group covariance property defining a phase space representation for the group $\SU$.

Now consider the discrete representation on sphere defined by Heiss and Weigert.  Now the phase space $\Lambda$ is a subset of points on the sphere which form a valid constellation.  Consider the frame $\mathcal F^{\textrm{constellation}}:=\{F^{\textrm{constellation}}(\nu)\}$, where
\begin{equation*}
F^{\textrm{constellation}}(\nu)=  \triangle_\nu,
\end{equation*}
where $  \triangle_\nu$ is a kernel satisfying the postulates \eqref{discrete_kernel_1}-\eqref{discrete_kernel_4}.
The quasi-probability function $\mu_\rho^{\textrm{constellation}}$ is a frame representation given by the frame $\mathcal F^{\textrm{constellation}}$.
As was the case for the other discrete representation in the previous examples, the frame operator of $\mathcal F^{\textrm{constellation}}$ is
$ S=d^{-1}\tilde \id$.  The unique dual frame of $\mathcal F^{\textrm{constellation}}$ is given
by $ S^{-1}\mathcal F^{\textrm{constellation}}$, where here $ S^{-1}=d\tilde \id$. Thus, the
dual frame to $\mathcal F^{\textrm{constellation}}$ is what Heiss and Weigert call a dual kernel.  Again, from Equation \eqref{discrete_kernel_4}, this representation satisfies definition of a phase space representation.

\subsubsection{Finite fields discrete phase space representation\label{sec:Example_Fields}}

Let $d$ be a power of a prime number.  Here $\Lambda=\field F_d\times\field F_d$.  Consider the frame $\mathcal F^{\textrm{field}}=\{F^{\textrm{field}}(q,p)\}$, where
\begin{equation*}
F^{\textrm{field}}(q,p)=\frac{1}{d}\left(\sum_{(q,p)\in\lambda}   Q(\lambda)-  \id\right),\label{fieldframe}
\end{equation*}
where $Q$ is a quantum net.  The quasi-probability function $\mu_\rho^{\textrm{field}}$ is a frame representation given by the frame $\mathcal F^{\textrm{field}}$.
The frame operator of $\mathcal F^{\textrm{field}}$ is
$ S=d^{-1}\tilde \id$.  The unique dual frame of $\mathcal F^{\textrm{field}}$ is given
by $ S^{-1}\mathcal F^{\textrm{field}}$, where here $ S^{-1}=d\tilde \id$. The
dual frame to $\mathcal F^{\textrm{field}}$ is a set of phase point
operators.

Equation \eqref{transcov} shows this particular representation is constructed to be translationally covariant and is thus a phase space representation.

\section{Frame representations of quantum theory\label{sec:quasi-probabilityrepsof}}

Quantum theory is an operational theory where each preparation is
associated with a density operator $\rho\in\den$. This association
is not required to be injective; different preparations may lead to
the same density operator. However, the mapping is assumed to be
surjective; there exists a preparation which leads to each density
operator.  Similarly, each measurement procedure and outcome $k$ is
associated with an effect $E_k\in\eff$. Again, this mapping need not
be injective but it is surjective. Quantum theory prescribes the
probability of outcome $k$ via the Born rule $\Pr(k)=\Tr(\rho
E_{k})$ . Since the set of outcomes is mutually exclusive and
exhaustive, each measurement procedure together will all the outcomes $\{E_{k}\}$ forms a positive operator valued measure (POVM).

Table \ref{table:qp} shows that most proposed quasi-probability
representations are representations of quantum states alone.  In
Sections \ref{sec:deformed_prob_frame} and
\ref{sec:quasi_prob_frame} we show that there are two approaches
within the frame formalism to lift any representation of states to a
fully autonomous representation of finite dimensional quantum
theory.  In Section \ref{sec:consis_cond}, a set of internal
consistency conditions for each of the two approaches is given that
allows one to formulate quantum theory independently of the standard
operator theoretic formalism. A short detour is taken in Section
\ref{sec:Transformations} to show that an operational formulation of
quantum theory including \emph{transformations} can be accommodated
within the scope of the frame formalism.  Finally in Section
\ref{sec:Example_SICPOVM} a novel quasi-probability representation
based on \emph{SIC-POVMs} is presented which relates the frame
formalism to a recent research topic in quantum information.

\subsection{Deformed probability representations of quantum theory\label{sec:deformed_prob_frame}}

The first frame representation approach consists of mapping both states and measurements to $L^2({\Lambda} )$ via a particular choice of frame $\mathcal F$.  Formally, we have the following definition:
\begin{definition}\label{def:deformed frame representation}
Suppose that the frame $\{F(\lambda)\}$ consists of positive operators which satisfies $\int_\Lambda d\lambda\; F(\lambda)=\id$ (a normalization condition).  Together, the mappings
\begin{align*}
\rho\mapsto\ip{F(\lambda)}{\rho}\\
E\mapsto\ip{F(\lambda)}{E},
\end{align*}
for all $\rho\in\den$ and $E\in\eff$, are called a \emph{deformed probability representation of quantum theory}.
\end{definition}

The reason for the qualifier \emph{deformed} is apparent from the following proposition:
\begin{proposition}\label{prop:deformed}
A deformed probability representation of quantum theory is a pair of mappings (call them $\mu_\rho:=\ip{F}{\rho}$ and $\xi_E=\ip{F}{E}$) which satisfy, for all $\lambda\in\Lambda$, all $\rho\in\den$ and all $E\in\eff$,
\begin{enumerate}[(a)]
\item $\mu_\rho(\lambda)\geq0$ and $\xi_E(\lambda)\in[0,1]$;
\item  $\int_\Lambda d\lambda\; \mu_\rho(\lambda)=1$; and,
\item $\Tr(\rho E)=\int_{\Lambda^2} d\lambda d\gamma\;  \mu_\rho(\lambda) \xi_E(\gamma) \ip{D(\lambda)}{D(\gamma)}$,
\end{enumerate}
where $\{D(\lambda)\}$ is a frame dual to $\{F(\lambda)\}$.
\end{proposition}

Property (c) is a \emph{deformed} law of total probability.  It can
be thought of as a deformation of the usual law of total
probability.  Recall from Lemma \ref{lemma_frame} that all
quasi-probability representations of states are frame
representations.  Given a quasi-probability representation (of
states), one can identify the unique frame which gives rise to it.
Then, using that frame to represent the measurement operators, one
obtains a deformed probability representation of quantum theory.

\subsection{Frame representations of quantum theory\label{sec:quasi_prob_frame}}

Notice that the deformed probability calculus (Proposition \ref{prop:deformed}(c)) can be written
\begin{equation}
\Tr(\rho E)=\int_\Lambda d\lambda\; \mu_\rho(\lambda)\xi'_{E}(\lambda),\label{lawtotalprob}
\end{equation}
where
\begin{equation}
 \xi'_{E}(\lambda)=\int_\Lambda d\gamma\; \xi_{E}( \gamma)\ip{ D(\lambda)}{ D(\gamma)}.\label{E'}
\end{equation}
Recall that $\xi_E$ is the frame representation of $ E$ for the frame $\mathcal F$.  Hence $\xi'_{E}$ can be identified as the frame representation of $ E$ using a frame $\mathcal D$ that is dual to $\mathcal F$.   The second frame representation approach consists of mapping states to $L^2(\Lambda )$ via a particular choice of frame $\mathcal F$ and measurements to $L^2(\Lambda)$ via a frame $\mathcal D$ that is dual to $\mathcal F$.  Formally, we have the following definition:
\begin{definition}\label{def:frame representation of quantum theory}
Suppose that the frames $\{F(\lambda)\}$ and $\{D(\lambda)\}$ are dual and $\int_\Lambda d\lambda\; F(\lambda)=\id$ (normalization).  Together, the mappings
\begin{align*}
\rho\mapsto\ip{F(\lambda)}{\rho}\\
E\mapsto\ip{D(\lambda)}{E},
\end{align*}
for all $\rho\in\den$ and $E\in\eff$, are called a \emph{frame representation of quantum theory}.
\end{definition}

A frame representation of quantum theory satisfies properties
similar to those of a deformed probability representation.
\begin{proposition}\label{prop:qp rep quant theory}
A frame representation of quantum theory is a pair of mappings (call them $\mu_\rho:=\ip{F}{\rho}$ and $\xi'_E:=\ip{D}{E}$)
which satisfy, for all $\lambda\in\Lambda$, all $\rho\in\den$ and
all $E\in\eff$,
\begin{enumerate}[(a)]
\item $\mu_\rho(\lambda)\in\mathbb R$ and $\xi'_E(\lambda)\in\mathbb R$;
\item $\int_\Lambda d\lambda\; \mu_\rho(\lambda)=1$; and,
\item $\Tr(\rho E)=\int_{\Lambda} d\lambda\;  \mu_\rho(\lambda) \xi'_E(\lambda)$.
\end{enumerate}
\end{proposition}

Property (c) is formally identical to the usual law of total
probability although the functions in the integrand are not presumed
to be non-negative.  Indeed, in \cite{FE08} it was shown that a pair
of dual frames cannot both consist solely of positive operators.

Recall from Lemma \ref{lemma_frame} that all quasi-probability
representations of states are frame representations.  Given a
quasi-probability representation of states, one can identify the
unique frame which gives rise to it.  Then, using a \emph{dual
frame} to represent the measurement operators, one obtains a frame
representation of quantum theory.

\subsection{Internal consistency conditions\label{sec:consis_cond}}

Consider first the deformed probability representations.  Of course, for a particular choice of frame, not every function in $L^2(\Lambda)$ will correspond to a valid quantum state or effect.  Here a set of \emph{internal} conditions is provided, independent of the standard axioms of quantum theory, which characterize the valid functions in $L^2(\Lambda)$.  The conditions can be found by noting that the frame
representation Equation \eqref{def_frame_rep} is an isometric and algebraic
isomorphism from $\herm$ to $L^2(\Lambda)$ equipped with
inner product
\begin{equation*}
\ip{\mu_1}{\mu_2}_\tD:=\int_{\Lambda^2} d\lambda d\gamma\;{\mu_1(\lambda)}\mu_2(\gamma)\tD(\lambda,\gamma),\label{def_ipGamma}
\end{equation*}
where $\tD(\lambda,\gamma):=\ip{ D(\lambda)}{ D(\gamma)}$, and algebraic multiplication
\begin{equation*}\label{def_starprod}
(\mu_1\star_\fF \mu_2)(\lambda):=\int_{\Lambda^2} d\gamma d\eta\;\mu_1(\gamma)
\mu_2(\eta) \fF(\lambda,\gamma,\eta),
\end{equation*}
where $\fF(\lambda,\gamma,\eta)=\ip{ F(\lambda)}{ D(\gamma)
 D(\eta)}$.

Now the condition for a function in $L^2(\Lambda)$ to be a valid state or effect can be stated.  A \emph{pure state} is a function
$\mu_\pure\in L^2(\Lambda)$ satisfying $\mu_\pure\star_\fF
\mu_\pure=\mu_\pure$.  A general \emph{state} is a function
$\mu\in L^2(\Lambda)$ satisfying
$\ip{\mu}{\mu_\pure}_\tD\geq0$ for all pure states
and $\int_\Lambda d\lambda\;
\mu(\lambda)=1$.  A \emph{measurement} is represented by
a set $\{\xi_k\in L^2(\Lambda)\}$ of \emph{effects} which satisfies
$\ip{\xi_k}{\mu_\pure}_\tD\geq0$ for all pure states
and for which
$\sum_{k}\xi_k=\xi_\id,$ where $\xi_\id$ is the identity element in $L^2(\Lambda)$ with
respect to the algebra defined by $\star_\fF$.  That is, $\xi_\id$ is the unique element satisfying $\xi_\id\star_\fF \mu= \mu\star_\fF \xi_\id$ for all $\mu\in L^2(\Lambda)$.

Now consider the frame representations.  Again for this approach, states and measurements in $L^2(\Lambda )$ must meet certain criteria to be valid.  The conditions are similar to those in the deformed probability representation.  Indeed the pure states and general states are equivalently characterized.  However, a measurement is now represented by a set $\{\xi'_k\in L^2(\Lambda)\}$ which satisfies
$\ip{\xi'_k}{\mu_\pure}\geq0$ (now the usual pointwise inner product) for all pure states
and for which
$\sum_{k}\xi'_k=\xi'_\id,$ where $\xi'_\id$ is the identity element in $L^2(\Lambda)$ with
respect to the algebra defined by $\star_{\mathfrak E}$ (which is defined in the same way as $\star_\fF$ with the roles of the frame and its dual reversed).

\subsection{Transformations\label{sec:Transformations}}

A transformation is a superoperator (an operator acting on operators) $\Phi:\den\to\den$.  Operationally, an experiment consists of preparations followed by transformations and ending in a measurement.  Note that, in a purely mathematical sense, the transformations are somewhat redundant as they could be bundled with either the preparations (to make new preparations) or measurements (to make new measurements).

A completely positive (CP) map is a linear superoperator $\Phi$ satisfying
\begin{equation*}
\Tr[(\Phi\otimes  \id)  \rho]\geq 0, \label{defCP}
\end{equation*}
for every pure state $ \rho$ on an extended system of arbitrary
finite dimension. If in addition, the CP map satisfies
$\Tr(\Phi( \rho))=\Tr( \rho)$, it is called a completely
positive trace-preserving (CPTP) map.  The CPTP maps are
transformations which are physically admissible.

In classical theories, transitions in probability are represented by matrices called \emph{stochastic matrices}.  It is natural to attempt a similar representation of transitions of quantum states here.  Matrix representation are typical in quantum theory.  A linear operator $A$ is usually mapped to a matrix with entries $a_{ij}$ given by $a_{ij}=\ip{\phi_i}{A\phi_j}$ where $\{\phi_i\}$ is an orthonormal basis for $\Hil$.  Then the action of the operator is representation as the usual matrix multiplication.  However, a slightly modified approach is required here when using frames (which reduces to usual matrix representations when the frame and an orthonormal basis coincide).

Let $\mu_\rho(\lambda)$ be a frame representation of a density operator $\rho$ for a frame $\mathcal F$.  Let $\mathcal D$ be a dual frame of $\mathcal F$ and consider the action of a superoperator
\begin{equation}\label{supop}
\Phi \rho= \Phi \int_\Lambda d\lambda\;  \mu_\rho(\lambda) D(\lambda).
\end{equation}
The frame representation of $\Phi \rho$ is $\mu_{\Phi \rho}(\gamma)= \ip{ F(\gamma)}{\Phi \rho}$.  As was the case for including measurements into a frame representations, two approaches can be identified for including the transformations into the frame representation formalism.

The first approach follows directly from Equation \eqref{supop}
\begin{equation}\label{superop_qp}
\mu_{\Phi \rho}(\gamma)=\int_\Lambda d\lambda\;  \Phi^{\textrm{qp}}(\gamma,\lambda) \mu_\rho(\lambda),
\end{equation}
where $\Phi^{\textrm{qp}}(\gamma,\lambda)=\ip{ F(\gamma)}{\Phi D(\lambda)}$ (``qp'' is a label meant to abbreviate ``quasi-probability'').  Notice that Equation \eqref{superop_qp} is just the usual (perhaps infinite dimensional) matrix multiplication rule.  It is the same rule for transitioning probability distributions via stochastic matrices in classical theories.  However, as opposed to stochastic matrices, $\Phi^{\textrm{qp}}$ could have negative entries.

Alternatively, consider the intermediate step
\begin{equation}\label{dual_rep}
 D(\lambda)=\int_\Lambda d\lambda\;  \ip{ D(\eta)}{ D(\lambda)} F(\eta).
\end{equation}
Then Equation \eqref{supop} becomes
\begin{equation}\label{superop_def}
\mu_\rho^\Phi(\gamma)=\int_{\Lambda^2} d\gamma d\eta\; \Phi^{\textrm{def}}(\gamma,\eta) \tD(\eta,\lambda) \mu_\rho(\lambda),
\end{equation}
where $\Phi^{\textrm{def}}(\gamma,\eta)=\ip{ F(\gamma)}{\Phi F(\eta)}$ (``def'' is a label meant to abbreviate ``deformed probability'').  This second approach is analogous to the deformed probability representation of Section \ref{sec:deformed_prob_frame}.

\subsection{Example: SIC-POVM representation\label{sec:Example_SICPOVM}}

In \cite{RBSC04}, the authors conjecture\footnote{Apparently this
was conjectured earlier by Zauner in a Ph.D. thesis not available in
english.  See \url{http://www.imaph.tu-bs.de/qi/problems/23.html}.} that
the set
$\{\phi_\alpha\in\mathcal H:\alpha\in\mathbb Z_d\times\mathbb Z_d\}=\{U_{(p,q)}\phi:
(p,q)\in\mathbb Z_d\times\mathbb Z_d\}$ for some $\phi\in\mathcal H$ and
\begin{equation}
U_{(p,q)}=\omega^{\frac{pq}{2}}X^pZ^q\label{Weylopts}
\end{equation}
forms a \emph{symmetric informationally complete positive operator valued measure} (SIC-POVM).  The defining condition of a SIC-POVM is
\begin{equation}
|\ip{\phi_\alpha}{\phi_\beta}|^2=\frac{\delta_{\alpha\beta}d+1}{d+1}.\label{SICPOVMdef}
\end{equation}
The set is called symmetric since the vectors have equal overlap.  The POVM is formed by taking the projectors onto the one-dimensional subspaces spanned by the vectors.  It is informationally complete since these $d^2$ projectors span $\herm$.

As of writing, it is an open question whether SIC-POVMs exist in
every dimension. However, there is numerical evidence for there
existence for every dimension up to $d=45$ \cite{RBSC04} and
analytic construction for a small number of dimensions.


Suppose then that for any dimension $d$, a SIC-POVM exists.  Notice that a SIC-POVM forms a frame.  Explicitly, let
\begin{equation*}
\mathcal
F=\left\{ F_\alpha:=\frac{1}{d}\phi_\alpha\phi_\alpha^\ast:\alpha\in\mathbb Z_d\times\mathbb Z_d\right\}\label{SICPOVMframe}
\end{equation*}
denote this frame.  From the definition of the SIC-POVM, Equation \eqref{SICPOVMdef},
\begin{equation*}
\mathtt F_{\alpha\beta}:=\ip{ F_\alpha}{ F_\beta}=\frac{\delta_{\alpha\beta}d+1}{d^2(d+1)}.\label{SICPOVM_Gram}
\end{equation*}
Since the frame forms a basis, the dual frame is unique and thus the inverse frame operator must satisfy
\begin{equation*}
\ip{ F_\alpha}{ S^{-1} F_\beta}=\delta_{\alpha\beta}.\label{SICPOVM_invframeop}
\end{equation*}
By inspection
\begin{equation*}
 D_\beta= S^{-1} F_\beta=d(d+1) F_\beta- \id.\label{SICPOVM_recipframe}
\end{equation*}
Representing a quantum state via the frame or canonical dual yields the neat reconstruction formulae
\begin{align}
\rho&=\sum_\alpha(d(d+1)\mu_\rho(\alpha)-1) F_\alpha,\label{SICPOVM_reconstruct1}\\
\rho&=\sum_\alpha\mu_\rho(\alpha)(d(d+1) F_\alpha- \id),\label{SICPOVM_reconstruct2}
\end{align}
where $\mu_\rho(\alpha):=\ip{ F_\alpha}{ \rho}$ is the frame representation of $ \rho$.

Equation \eqref{SICPOVM_reconstruct1} was given in \cite{ADF07}.  This equation fits naturally into the deformed probability representation formalism discussed in Section \ref{sec:deformed_prob_frame}.  Notice that the dual frame satisfies
\begin{align*}
\tD_{\alpha\beta}&=\ip{ D_\alpha}{ D_\beta}\\
&=\ip{d(d+1) F_\alpha- \id}{d(d+1) F_\beta- \id}\\
&=d^2(d+1)^2\tF_{\alpha\beta}-2(d+1)+d\\
&=d(d+1)\delta_{\alpha\beta}-1.
\end{align*}
If an arbitrary measurement $\{ E_k\}$ is also represented via the SIC-POVM frame as $\xi_{E_k}(\beta):=\ip{ F_\beta}{ E_k}$, then Equation \eqref{SICPOVM_reconstruct1} is identical to the deformed law of total probability
\[
\Pr(k)=\sum_{\alpha\beta} \mu_\rho(\alpha) \xi_{E_k}(\beta)  \tD_{\alpha\beta}.
\]
Equation \eqref{SICPOVM_reconstruct2} fits more naturally into the frame representation formalism discussed in Section \ref{sec:quasi_prob_frame}. If an arbitrary measurement $\{ E_k\}$ is represented via the canonical dual to the SIC-POVM frame as $\xi'_{E_k}(\alpha):=\ip{ D_\alpha}{ E_k}$, then Equation \eqref{SICPOVM_reconstruct2} is identical to the usual law of total probability
\[
\Pr(k)=\sum_{\alpha} \mu_\rho(\alpha) \xi'_{E_k}(\alpha).
\]

Since the SIC-POVM frame is made of projectors, the frame representation of the density operator is a true probability distribution.
However, the dual frame operators are not positive.  Thus in a SIC-POVM frame representation, the measurement objects (meant to represent conditional probabilities) contain negative values while the states are represented by true probabilities.  The appearance of negativity is a necessary feature of frame representations, as was shown in \cite{FE08}.

\section{Non-classicality of quantum theory\label{sec:necessity of negativity}}
In this section we prove the impossibility of representing quantum theory classically via a mapping from the quantum operators to classical probability functions.

\subsection{Classical representations of quantum theory\label{sec:classicalrepsquantum}}

In classical probability, we postulate that a physical system has a
set of properties mathematically represented by a measure space
$\Lambda$.  The ontological states are represented by the Dirac
measures which are the extreme points of the set of all probability
measures on $\Lambda$.  The probability measures are the epistemic
states representing our ignorance of the ontic state of the system.
To measure the properties of the system we partition the space
$\Lambda$ into disjoint subsets $\{\triangle_k\}$.  The probability
of the system to have properties in $\triangle_k$ (we will call this
``outcome $k$'') is
\begin{equation*}
\Pr(k|\mu)=\int_{\triangle_k}d\lambda\;\mu(\lambda)=\int_\Lambda d\lambda\; \mu(\lambda)\chi_{k}(\lambda),
\end{equation*}
where $\chi_k(\lambda)\in\{0,1\}$ is an idempotent indicator
function associated with $\triangle_k$.  The measurement is
equivalently specified by the set $\{\chi_k(\lambda)\}$, which is
interpreted as the conditional probability of outcome $k$ given the
systems is known to have the properties $\lambda$.  A measurement of
this type is deterministic; it reveals with certainty the properties
of the system.  Consider now an \emph{indeterministic} measurement
specified by the conditional probabilities
$\{\xi_k(\lambda)\in[0,1]\}$.  Each of these can be decomposed as a
convex combination of idempotent indicator functions.  For this
reason, idempotent indicator functions are emphasized as being
\emph{sharp} within the set of more general measurement functions.
Put concisely, the indicator functions form a convex set with the
sharp indicator functions as the extreme points.

Consider the problem of proving whether or not it is possible to
construct a representation of quantum theory that satisfies these
classical postulates. That is, we are interested if there exists a
space $\Lambda$ such that for every density operator there is a unique
probability measure and every effect there is an unique indicator function
on this space which reproduce the Born rule via the law of total
probability. We can formalize this idea into the following
definition.

\begin{definition}\label{def:classical}
A \emph{classical representation of quantum theory} is a pair of mappings $\mu$ and $\xi$ whose domains are $\den$ and $\eff$, respectively, and whose co-domains are a convex set of probability functions.  These mappings satisfy, for all $\rho\in\den$ and all $E\in\eff$,
\begin{enumerate}[(a)]
\item $\mu_\rho(\lambda)\geq0$ and $\xi_E(\lambda)\in[0,1]$;
\item $\int_\Lambda d\lambda\; \mu_\rho(\lambda)=1$; and,
\item $\Tr(\rho E)=\int_\Lambda d\lambda\;\mu_\rho(\lambda)\xi_E(\lambda).$
\end{enumerate}
\end{definition}

We submit Definition \ref{def:classical} as the most natural set of requirements for what can reasonably be called a ``classical representation'' of quantum theory.

\subsection{Quasi-probability representations of quantum theory\label{sec:affinerepsquantum}}

In light of the above considerations, the most natural approach to obtaining a set of quasi-probability representations is to take the definition of a  classical representation and simply relax the requirement of non-negativity. This motivates defining the following set of generalized
representations.

\begin{definition}\label{def:qp rep quant theory}
A \emph{quasi-probability representation of quantum theory} is a
pair of mappings $\mu$ and $\xi$ whose domains are $\den$ and
$\eff$, respectively, and whose co-domains are a convex set of quasi-probability functions (real valued functions).  These mappings satisfy, for all $\rho\in\den$
and all $E\in\eff$,
\begin{enumerate}[(a)]
\item $\mu_\rho(\lambda)\in\mathbb R$ and $\xi_E(\lambda)\in\mathbb R$;
\item $\int_\Lambda d\lambda\; \mu_\rho(\lambda)=1$; and,
\item $\Tr(\rho E)=\int_\Lambda d\lambda\;\mu_\rho(\lambda)\xi_E(\lambda) .$
\end{enumerate}
\end{definition}

It should be obvious that a non-negative
quasi-probability representation is just a classical representation.  Now we show that a consequence of these requirements is that a quasi-probability satisfies a very important property which will be used later on.

\begin{lemma}\label{prop:classsical rep is affine}
The mappings in a quasi-probability representation of quantum theory are affine\footnote{Note that a few authors use the terminology ``convex-linear'' to mean affine, the latter being a well understood and precise mathematical term: a mapping which preserves convex combinations.}.
\end{lemma}
\begin{proof}
First note from property (c) that we have
\begin{equation}\label{classical rep non-contextuality}
\int_\Lambda d\lambda\;\mu_\rho(\lambda)\xi_E(\lambda)=\int_\Lambda d\lambda\;\mu_\sigma(\lambda)\xi_E(\lambda)\Rightarrow \Tr(\rho E)=\Tr(\sigma E)\Rightarrow \rho=\sigma \Rightarrow \mu_\rho(\lambda)=\mu_\sigma(\lambda).
\end{equation}
Now consider the density operator $\rho = p \sigma_1 +(1-p)\sigma_2$, $0\leq p\leq 1$.  Multiplying by an arbitrary effect $E$ and taking the trace we have $\Tr(\rho E) = p \Tr(\sigma_1  E) +(1-p)\Tr (\sigma_2  E)$.  By property (c) and Eq. \eqref{classical rep non-contextuality} we have
\[
\int_\Lambda d\lambda\;\mu_\rho(\lambda)\xi_E(\lambda)=\int_\Lambda d\lambda\;\left(p\mu_{\sigma_1}(\lambda)+(1-p)\mu_{\sigma_2}(\lambda)\right)\xi_E(\lambda)\Rightarrow \mu_\rho(\lambda)=p\mu_{\sigma_1}(\lambda)+(1-p)\mu_{\sigma_2}(\lambda).
\]
Hence, $\mu$ is affine.  The proof that $\xi$ is affine is identical.
\end{proof}

Comparing Definition \ref{def:qp rep quant theory} to Proposition \ref{prop:qp rep quant theory} it would seem that a quasi-probability representation is equivalent to an frame representation, an vice versa.  It is important to note however, via Definition \ref{def:frame representation of quantum
theory}, that a frame representation of quantum theory makes explicit reference to frames.   Whereas, a quasi-probability representation of quantum theory is defined without any reference to frames.  However, we will see in the next section that the two are indeed equivalent.

\subsection{Equivalence of frame representations and quasi-probability representations\label{sec:equivaffineqp}}

In \cite{FE08} it was shown that there does not exist a frame of
positive operators that is dual to another frame of positive
operators. This establishes that frame representations of quantum
theory, as described in Section \ref{sec:quasi_prob_frame}, must
poses negativity. However, this proof does not rule out the
possibility of \emph{non-negative} quasi-probability representations (or, equivalently,
classical representations) as defined above.  We now close this
loophole and thereby extend the significance of the no-go theorem in
\cite{FE08} via the following lemma which shows that the set of
quasi-probability representations and the set of frame representations are
equivalent.

\begin{lemma}\label{lemma:quasi_to_frame}
A quasi-probability representation of quantum theory (Definition
\ref{def:qp rep quant theory}) is equivalent to a frame
representation of quantum theory (Definition \ref{def:frame
representation of quantum theory}).
\end{lemma}
\begin{proof}
A frame representation clearly satisfies Definition \ref{def:qp rep
quant theory}.  Suppose then that $\mu$ and $\xi$ form a quasi-probability representations of quantum theory.  First consider
$\mu$.  According the procedure described in \cite{Bus03}, this
mapping can be extended to a linear map on $\herm$.  Then the Riesz
representation theorem implies that there exists a unique operator
valued function $\mathcal{F} :=\{F(\lambda)\}$ such that
$\mu_A=\ip{A}{F}$.  To show $\mu_A\in L^2(\Lambda)$ we need to show
that $\int_\Lambda d\lambda\; |\mu_A(\lambda)|^2$ is finite.  Note
that
\begin{align*}
\int_\Lambda d\lambda\; \abs{\mu_A(\lambda)}^2=\int_\Lambda d\lambda\;
\abs{\ip{A}{F(\lambda)}}^2=\ip{A}{ S A},
\end{align*}
where $ S$ is the frame operator of $\mathcal F$.  Since the
range of the frame operator is another Hermitian operator and
$\herm$ is finite dimensional, $\ip{A}{ S A}<\infty$.  To show
that $\mu$ is injective we suppose $\mu_A=0$ and show $A=0$.  From
Property \ref{def:qp rep quant theory}(c)
\[
\Tr(AE)=\int_\Lambda d\lambda \;\mu_A(\lambda)\xi_E(\lambda)=0.
\]
Since $E$ is arbitrary and $\eff$ spans $\herm$, this immediately
implies $A=0$. Lemma \ref{lemma_frame} implies  $\mathcal F$ is a
frame.  The same logic implies $\xi$ is a frame representation for
some unique frame $\mathcal D:=\{D(\lambda)\}$.  To show that
together these form an quasi-probability representation,
we need only show that $\mathcal D$ is dual to $\mathcal F$.
Property \ref{def:qp rep quant theory}(c) implies
\[
\Tr(AB)=\int_\Lambda d\lambda\;\ip{F(\lambda)}{A}\ip{D(\lambda)}{B},
\]
which is true for all $B\in\herm$.  Thus
\[
A=\int_\Lambda d\lambda\;\ip{F(\lambda)}{A}D(\lambda).
\]
By Definition \ref{def:dual}, $\mathcal D$ is dual to $\mathcal F$.
\end{proof}

\subsection{Non-classicality of quantum theory\label{sec:nonclass}}

Using Lemma \ref{lemma:quasi_to_frame} in combination with the negativity theorem of \cite{FE08} it is easy to prove the following.

\begin{theorem}\label{theorem_negativity}
A classical representation of quantum theory (Definition \ref{def:classical}) does not exist.  Or, equivalently, a non-negative quasi-probability representation (Definition \ref{def:qp rep quant theory}) does not exist.
\end{theorem}
\begin{proof}
Lemma \ref{lemma:quasi_to_frame} establishes that a classical representation is a quasi-probability representation which requires that the dual of a positive frame be positive.  Such a pair of frames does not exist as proven in \cite{FE08}.  For the reader unfamiliar with frames this fact can also be proven directly, without making use of frames, as follows.

It is much easier to see the contradiction when $\Lambda$ is assumed to be finite.  It follows from \cite{Bus03} (see also \cite{CFMR04}) that $\mu_\rho(\lambda)=\ip{\rho}{F(\lambda)}$, where $F$ is some effect-valued function.  Similarly, $\xi_E(\lambda)=\ip{E}{D(\lambda)}$, where $D$ is state-valued function.  That is, each $D(\lambda)$ is a density operator.  From Property \ref{def:classical}(c) it follows that
\begin{equation}\label{nonclassproof}
\rho=\sum_\lambda \ip{\rho}{F(\lambda)}D(\lambda).
\end{equation}
This is true for all $\rho$ including the rank-1 projectors: the extreme points of $\den$.   Equation \eqref{nonclassproof} is a convex combination.  The fact that extreme point have only the trivial decomposition implies $\mu$ maps the extreme of $\den$ to Dirac distributions: the extreme points of probability distributions\footnote{Spekkens \cite{Spe08} has also made these observations but follows a different route from here on to obtain a contradiction.  See Section \ref{chap:contextuality} for details of the connection to Reference \cite{Spe08}.}.  There are only finitely many extreme points in the simplex of probability distributions over a finite space.  There are infinitely many extreme points in $\den$.   Property \ref{def:classical}(c) also implies $\mu$ must be injective\footnote{The injectivity of this map can be proven from Property\ref{def:classical}(c) as in the proof of Lemma \ref{lemma:quasi_to_frame}.}.  This is clearly impossible and so all hope of a classical representation must rest on the set $\Lambda$ having the same cardinality of $\den$\footnote{Compare this to the ``Ontological excess baggage theorem'' of Hardy \cite{Har04}.  Also note that Busch, Hellwig and Stulpe have come to the same conclusion for finite representations: quantum states can be mapped to probability vectors and quantum observables to classical random variables at the expense of quantum effects being mapped to non-classical (i.e. negative valued) objects \cite{BHS93}.}.

Now assume $\Lambda$ is infinite.  Property \ref{def:classical}(c) still requires $\mu$ must map the rank-1 projectors to the Dirac measures.  Recall for a finite dimensional simplex that each vector has a unique decomposition into extreme points.  The same is true here for the probability measures on $\Lambda$; each measure has a unique decomposition into Dirac measures \cite{Cho69}.  Since, in general, $\rho\in\den$ has infinitely many non-equivalent decompositions into rank-1 projectors, it is impossible for $\mu$ to be injective.
\end{proof}

Theorem \ref{theorem_negativity} establishes the necessity of negativity in quasi-probability representations of quantum theory.  In the next section another notion of non-classicality, \emph{contextuality}, will be compared with negativity.

\section{Connection with contextuality \label{chap:contextuality}}

Within the quantum formalism there are many notions of
non-classicality. In the previous section, negativity was proven to
be such a notion. In this section the idea of \emph{contextuality}
is considered as an alternative candidate.  In Section
\ref{sec:contextuality_traditional} the traditional notion of
contextuality is reviewed.  A generalization due to Spekkens is
presented in Section \ref{sec:contextuality_general}.  In Section
\ref{sec:negativity_and_contextuality} the somewhat misleading claim
in \cite{Spe08} that ``negativity and contextuality are equivalent''
is clarified.  Finally, a more general model of quantum theory, which allows both negativity and contextuality is considered in Section \ref{sec:generalmodels}.

\subsection{Traditional definition of contextuality \label{sec:contextuality_traditional}}

The traditional definition of contextuality evolved from a theorem
which appears in a paper by Kochen and Specker \cite{ks67}.  The
Kochen-Specker theorem concerns the standard quantum formalism:
physical systems are assigned states in a complex Hilbert space
$\Hil$ and measurements are made of observables represented by
Hermitian operators.  The theorem establishes a contradiction
between a set of plausible assumptions which together imply that
quantum systems possess a consistent set of pre-measurement values
for observable quantities. Let $\Hil$ be the Hilbert space
associated with a quantum system and $ A\in\herm$ be the operator
associated with an observable $A$.  The function $f_\psi(A)$
represents the value of the observable $A$ when the system is in
state $\psi$.  One assumption used to derive the contradiction is
that for any function $F$, $f_\psi(F(A))=F(f_\psi(A))$.  This is plausible because, for
example, we would expect that the value of $A^2$ could be obtained
in this way from the value of $A$.

Assuming that physical systems do possess values which can be
revealed via measurements, the Kochen-Specker theorem leads to the
following counterintuitive example \cite{Ish95}.  Suppose three
operators $ A$, $ B$, and $  C$ satisfy $[ A, B]=0=[ A,  C]$, but $[
B,  C]\neq0$.  Then the value of the observable $A$ will depend on
whether observable $B$ or $C$ is chosen to be measured as well. That
is, the value of $A$ depends on the \emph{context} of the
measurement.

What the Kochen-Specker theorem establishes then is the mathematical
framework of quantum theory does not allow for a
\emph{non-contextual} model for pre-measurement values.  This fact
is often expressed via the phrase ``quantum theory is contextual''.

\subsection{Generalized definition of contextuality \label{sec:contextuality_general}}
The original notion of contextuality only applies to measurements in
standard quantum theory.  It does not apply to general operational
theories, and, in particular, the mathematical model of open quantum
systems (i.e., consisting of mixed states, POVM measurements and
completely-positive maps). This problem was addressed by Spekkens in
Reference \cite{Spe05} as follows.

A general operational model (including classical and quantum theory)
specifies the probabilities $\Pr(k|\mathcal P,\mathcal M)$ for the
outcomes of a measurement procedure $\mathcal M$ given preparation
procedure $\mathcal P$. Each $\mathcal P$ belongs to an equivalence
class $e(\mathcal P)$ in which any two preparations, $\mathcal P$
and $\mathcal P'$ are equivalent if $\Pr(k|\mathcal P,\mathcal
M)=\Pr(k|\mathcal P',\mathcal M)$ for all $\mathcal M$. Each
$\mathcal M$ defines an equivalence class is a similar manner. The
features of an experimental configuration which are not specified by
the equivalence class of the procedure are called the \emph{context}
of the experiment.

One may supplement the operational theory with an ontic state space
$\Lambda$.  Then the preparation procedures become probability
densities $\mu_{\mathcal P}(\lambda)$ while the measurement
procedures become conditional probabilities $\xi_{\mathcal
M,k}(\lambda)$.  The probabilities of the outcomes of the
measurements is required to satisfy the law of total probability
\begin{equation}\label{ont_law_total_prob}
\Pr(k|\mathcal P,\mathcal M)=\int_{\Lambda} d\lambda\; \mu_\mathcal P (\lambda) \xi_{\mathcal M,k}(\lambda)
\end{equation}
Such a supplemented operation model is called an ontological model.
The ontological model is \emph{preparation non-contextual} if
\begin{equation}\label{prep_contextual}
\mu_\mathcal P (\lambda) = \mu_{e(\mathcal P)} (\lambda).
\end{equation}
That is, the representation of the preparation procedure is
independent of context.  Similarly the ontological model is
\emph{measurement non-contextual} if
\begin{equation}\label{meas_contextual}
\xi_{\mathcal M,k} (\lambda) = \xi_{e(\mathcal M),k} (\lambda).
\end{equation}
The terminology ``contextual'' is again shorthand for the inability
of an operational theory to admit a (preparation or measurement)
non-contextual ontological model. However, the term ``contextual''
is also used to describe specific ontological model which do not
satisfy Equations \eqref{prep_contextual} and
\eqref{meas_contextual} \cite{HR07}.  In Reference \cite{Spe05},
it was proven that quantum theory is both preparation and
measurement contextual.

Since the
standard quantum formalism is an instance of an operational model, Spekkens' notion of non-contextuality is a generalization of the
traditional notion initiated by Kochen and Specker.  Moreover, considering only measurements, one can see that Spekkens
generalizes the notion of \emph{non-contextuality} from outcomes of
individual measurements being independent of the measurement context
to \emph{probabilities} from outcomes of measurements being
independent of the measurement context.

\subsection{On the equivalence of non-negativity and non-contextuality \label{sec:negativity_and_contextuality}}
Recall that, in quantum theory, a preparation is specified by a
density operator $ \rho$ and a measurement outcome by an effect $
E$.  Thus a (preparation and measurement) non-contextual ontological
model of quantum theory would require
\begin{align*}
&\mu_\mathcal P(\lambda) = \mu_{ \rho}(\lambda),\\
&\xi_{\mathcal M}(\lambda)=\xi_{ E} (\lambda),
\end{align*}
and the law of total probability
\begin{align*}
\Tr( E \rho)=\int_{\Lambda}d\lambda\;\xi_{ E}(\lambda)\mu_{ \rho}(\lambda).
\end{align*}
Assuming the probabilities satisfy the usual normalization
conditions these conditions are equivalent to those proposed in
Definition \ref{def:classical} identifying a classical representation. Spekkens noticed this equivalence in \cite{Spe08}
and has therefore independently obtained a proof of negativity. Similarly,
our direct proof of the non-existence of a positive dual frame to a
positive frame gives a new independent proof that quantum theory satisfies Spekkens' generalized notion of
contextuality.

Note that the terminology ``negativity and contextuality are
equivalent'' used in \cite{Spe08}, is somewhat misleading.  A
quasi-probability representation that takes on negative values is
\emph{not} equivalent to some contextual ontological model.  Within the formalism
of quasi-probability representations of quantum theory, a classical representation (Definition \ref{def:classical}) is a special case (a non-negative one). A classical representation of
quantum theory is also a special case of a general operational
model (a non-contextual ontological model). The key
point is that non-negative quasi-probability representations and
non-contextual ontological models are the same thing, and one can establish the non-existence
of such a classical representation for quantum theory (Theorem
\ref{theorem_negativity}) starting from either formalism. In other
words, a classical representation does not exist but one can relax
the constraints on Definition \ref{def:classical} to achieve a
representation which may contain negativity or contextuality (or
both).

\subsection{General models for quantum theory \label{sec:generalmodels}}

A broader framework is to consider a very general class of models not requiring any classical features.  That is, suppose we define a general model as just a pair of \emph{relations} generalizing the (well-defined) \emph{mappings} $\mu$ and $\xi$ of Definition \ref{def:qp rep quant theory}.  In that case, quantum states and measurements are represented by functions over an ontic space
which allows for negativity, contextuality and a non-canonical (deformed)
probability calculus. Within such a framework we can consider the
following propositions: the model is non-negative (NN); the model
uses the law of total probability to reproduce the Born rule (LTP);
and the model is non-contextual (NC). This paper establishes that
the logical conjunction
\[
\mathrm{NN}\wedge\mathrm{LTP}\wedge\mathrm{NC}
\]
is false. We can deny NN to obtain a quasi-probability
representation. Alternatively,
we can deny LTP to obtain, for example, a deformed probability
representation (Section \ref{sec:deformed_prob_frame}).  We have
already seen how these two cases are intimately related.
Finally, one can instead deny NC while retaining LTP and NN,
which explicitly shows that negativity and contextuality are not
equivalent. An example of such a contextual model is discussed in Reference \cite{HR07}, section V A.

\section{Discussion\label{sec:conclusion}}

Although we conjecture that our no-go theorem will hold when $\Hil$
is infinite dimensional, it is clear the technique in the proof of Lemma
\ref{lemma:quasi_to_frame} will not suffice in that case.  The proof
makes explicit use of the finite dimensionality of $\Hil$.  However,
the alternate proof of Theorem \ref{theorem_negativity} does not
make explicit use of this fact. The key fact used in the alternate
proof is the non-simplex structure of the quantum state space
$\den$. This fact is true regardless of the dimension of $\Hil$.
However, it was also necessary that the quasi-probability
representation have the form $\mu_\rho=\ip{\rho}{F}$.  This form is guaranteed for infinite dimensions if we assume that the quasi-probability representation is \emph{bounded}: there exist $b>0$ such that for all $\rho\in\den$,
\begin{equation}\label{def:bounded}
\int d\lambda|\mu_\rho(\lambda)|^2\leq b\|\rho\|^2.
\end{equation}
If we demand boundedness on physical grounds, then our proofs hold for physically reasonable quasi-probability representations in infinite dimensional Hilbert spaces.

Finally, we would like to comment on potential applications of the
frame formalism and our analysis of the relationship between
non-negativity and non-contextuality. The question of ``What is
non-classical about quantum mechanics?'' has taken a more practical
and well-defined meaning within the context of quantum information
theory. On the one hand, the necessary and sufficient conditions on
the quantum resources required to outperform classical information
are not well understood \cite{KL98,Vid03}. On the other hand, the
rapid developments in quantum control in various experimental
settings, such as trapped ions, superconducting circuits, quantum
optics, and both liquid and solid-state NMR, and their application
as quantum information processors, has renewed interest in
understanding the extent to which specific experimental achievements
may be taken as evidence for truly coherent, quantum behaviour \cite{SC99,KLBR08}. It is our hope that the formalism developed in
this work will lead to more systematic and operationally meaningful
criteria for characterizing both of these issues.

\begin{acknowledgements}
The authors thank Bernard Bodmann, Matt Leifer, Etera Livine, Ryan Morris
and Rob Spekkens for helpful discussions. This work was supported by
NSERC and MITACS.
\end{acknowledgements}

\appendix

\section{Notation and definitions\label{sec:mathematical notations}}
Quantum theory makes use of complex Hilbert spaces.  If these spaces are finite dimensional, then they are equivalent to inner product spaces.  Unless otherwise noted, a Hilbert space $\Hil$ will be assumed to have dimension $d<\infty$ and (for $\psi,\phi\in\Hil$) its inner product will be denoted $\ip{\psi}{\phi}$.  The following list defines some specials sets of linear operators acting on $\Hil$.

\begin{enumerate}
\item An operator $A$ satisfying
\begin{equation}\label{def:hermitian}
\ip{A\psi}{\phi}=\ip{\psi}{A\phi},
\end{equation}
for all $\psi,\phi\in\Hil$, is called \emph{Hermitian}.  The set of all Hermitian operators is denoted $\herm$.
\item An operator $E$ satisfying
\begin{equation}\label{def:effect}
0\leq\ip{\psi}{E\psi}\leq1,
\end{equation}
for all $\psi\in\Hil$, is called an \emph{effect}.  The set of all effects is denoted $\eff$.
\item An effect $\rho$ satisfying
\begin{equation}\label{def:densityop}
\Tr(\rho)=1
\end{equation}
is called a \emph{density operator}.  The set of all density operators is denoted $\den$.
\item A set of effects $\{E_k\}$ which satisfy
\begin{equation}\label{def:povm_sum_rule}
\sum_k E_k=\id
\end{equation}
is called a POVM (positive operator valued measure).  The set of all POVMs is denoted $\povm$.
\item An effect $P$ satisfying
\begin{equation}\label{def:projector}
P^2=P
\end{equation}
is called a \emph{projector}.  The set of all projects is denoted $\proj$.
\item A set of projectors $\{P_k\}$, each of rank 1, which satisfy
\begin{equation}\label{def:povm_sum_rule}
\sum_k P_k=\id
\end{equation}
is called a PVM (projector valued measure).  The set of all PVMs is denoted $\pvm$.
\end{enumerate}

Note that from these definitions we have $\proj\subset\den\subset\eff\subset\herm$ and $\pvm\subset\povm$.  The set $\herm$ defines its own Hilbert space with inner product (for $A,B\in\herm$) $\ip{A}{B}:=\Tr(AB)$.  The dimension of $\herm$ is $d^2$.  A PVM contains exactly $d$ projectors which satisfy the \emph{orthogonality condition} $P_kP_j=P_k\delta_{kj}$, for any $k$ and $j$.

Here are some examples of important Hermitian operators used in this paper.  Consider the operator $Z$ whose spectrum is $\spec(Z)=\{\omega^k: k\in \field Z_d\}$, where $\omega^k=\w k$.  The
eigenvectors form a basis for $\Hil$ and are denoted
$\{\phi_k\}$. Consider also the operator defined by
$ X\phi_k=\phi_{k+1}$, where all arithmetic is modulo $d$.  Define $Y$ implicitly through $[ X, Z]=2i Y$.  The
operators $ Z,  X$ and $ Y$ are often called \emph{generalized
Pauli operators} since they are indeed the usual Pauli operators
when $d=2$.  The \emph{parity operator} is defined by $ P\phi_k=\phi_{-k}$.

\bibliographystyle{unsrt}
\bibliography{mybib}

\begin{thebibliography}{10}

\bibitem{Wig32}
Eugene Wigner.
\newblock On the quantum correction for thermodynamic equilibrium.
\newblock {\em Phys. Rev.}, 40:0749, 1932.

\bibitem{Lee94}
Hai-Woong Lee.
\newblock Theory and application of the qunatum phase-space distrbution
  functions.
\newblock {\em Phys. Rep.}, 259:147, 1994.

\bibitem{Bak58}
George~A. Baker.
\newblock Formulation of quantum mechanics based on the quasi-probability
  distribution induced on phase space.
\newblock {\em Phys. Rev.}, 109:2198, 1958.

\bibitem{Woo87}
William~K. Wootters.
\newblock A wigner-function formulation of finite-state quantum mechanics.
\newblock {\em Ann. Phy.}, 176:1, 1987.

\bibitem{GHW04}
Kathleen~S. Gibbons, Matthew~J. Hoffman, and William~K. Wootters.
\newblock Discete phase space based on finite fields.
\newblock {\em Phys. Rev. A}, 70:062101, 2004.
\newblock http://arxiv.org/abs/quant-ph/0401155v6.

\bibitem{Paz02}
Juan~Pablo Paz.
\newblock Discrete wigner functions and the phase-space representation of
  quantum teleportation.
\newblock {\em Phys. Rev. A}, 65:062311, 2002.
\newblock http://arxiv.org/abs/quant-ph/0204150v1.

\bibitem{LP03}
Cecilia~C. L\'{o}pez and Juan~Pablo Paz.
\newblock Phaes-space approach to the study of decoherence in quantum walks.
\newblock {\em Phys. Rev. A}, 68:0052305, 2003.
\newblock http://arxiv.org/abs/quant-ph/0308104v3.

\bibitem{MPS02}
Cesar Miquel, Juan~Pablo Paz, and Marcos Saraceno.
\newblock Quantum computers in phase space.
\newblock {\em Phys. Rev. A}, 65:062309, 2002.

\bibitem{Gal05}
Ernesto~F. Galvao.
\newblock Discrete wigner functions and quantum computational speedup.
\newblock {\em Phys. Rev. A}, 71:042302, 2005.
\newblock http://arxiv.org/abs/quant-ph/0405070v2.

\bibitem{CGG06}
Cecilia Cormick, Ernesto~F. Galvao, Daniel Gottesman, Juan~Pablo Paz, and
  Arthur~O. Pittenger.
\newblock Classicality in discrete wigner functions.
\newblock {\em Phys. Rev. A}, 73:012301, 2006.

\bibitem{GJ07}
D.~Gross and J.~Eisert.
\newblock Quantum margulis expanders, 2007.
\newblock http://arxiv.org/abs/quant-ph/0710.0651v1.

\bibitem{FE08}
Christopher Ferrie and Joseph Emerson.
\newblock Frame representations of quantum mechanics and the necessity of
  negativity in quasi-probability representations.
\newblock {\em J. Phys. A: Math. Theor.}, 41:352001, 2008.

\bibitem{Moy49}
Jose~E. Moyal.
\newblock Quantum mechanics as a statistical theory.
\newblock {\em Proc. Camb. Phil. Soc.}, 45:99, 1949.

\bibitem{BB87}
J.~Bertrand and P.~Bertrand.
\newblock A tomographic approach to wigner's function.
\newblock {\em Foundations of Phyiscs}, 17:397, 1987.

\bibitem{HOSW84}
M.~Hillery, R.F. O'Connell, M.O. Scully, and E.P. Wigner.
\newblock Distrubution functions in physics: fundamentals.
\newblock {\em Physics Reports}, 106:121, 1984.

\bibitem{CCSS87}
O.~Cohendet, Ph. Combe, M.~Sirugue, and M.~Sirugue-Collin.
\newblock A stochastic treatment of the dynamics of an integer spin.
\newblock {\em J. Phys. A: Math. Gen.}, 21:2875, 1987.

\bibitem{Leo95}
Ulf Leonhardt.
\newblock Quantum state tomography and discrete wigner functions.
\newblock {\em Phys. Rev. Lett.}, 74:4101, 1995.

\bibitem{Leo96}
Ulf Leonhardt.
\newblock Discrete wigner functions and quantum state tomography.
\newblock {\em Phys. Rev. A}, 53:2998, 1996.

\bibitem{HW00}
Stephan Heiss and Stefan Weigert.
\newblock Discrete moyal-type representations for a spin.
\newblock {\em Phys. Rev. A}, 63:012105, 2000.

\bibitem{Str57}
R.~L. Stratonovich.
\newblock {\em JETP}, 4:891, 1957.

\bibitem{Woo04}
William~K. Wootters.
\newblock Picturing qubits in phase space.
\newblock {\em IBM J. Rev. \& Dev.}, 48:99, 2004.

\bibitem{Vou04}
A.~Vourdas.
\newblock Quantum systems with finite hilbert space.
\newblock {\em Rep. Prog. Phys.}, 67:267, 2004.

\bibitem{Har01}
Lucien Hardy.
\newblock Quantum theory from five reasonable axioms, 2001.
\newblock http://arxiv.org/abs/quant-ph/0101012v4.

\bibitem{Hav03}
Timothy~F. Havel.
\newblock The real density matrix.
\newblock {\em Quantum Information Processing}, 1:511, 2003.
\newblock http://arxiv.org/abs/quant-ph/0302176v5.

\bibitem{RMG05}
M~Ruzzi, M.A. Marchiolli, and D.~Galetti.
\newblock Extended cahill-glauber formalism for finite-dimensional spaces: I.
  fundamentals.
\newblock {\em J. Phys. A: Math. Gen.}, 38:6239, 2005.

\bibitem{CEMMMS06}
S.~Chaturvedi, E.~Ercolessi, G.~Marmo, G.~Morandi, and R.~Simon.
\newblock Wigner-weyl correspondence in quantum mechanics for continuous and
  discrete systems - a dirac-inspired view.
\newblock {\em J. Phys. A: Math. Gen.}, 39:1405, 2006.

\bibitem{Gro06}
David Gross.
\newblock Hudson's theorem for finite-dimensional quantum systems.
\newblock {\em J. Math. Phys.}, 47:122107, 2006.
\newblock http://arxiv.org/abs/quant-ph/0602001v3.

\bibitem{Chr03}
Ole Christensen.
\newblock {\em Introduction to Frames and Riesz Bases}.
\newblock Birkh{\"a}user, Boston, 2003.

\bibitem{RBSC04}
Joseph~M. Renes, Robin Blume-Kohout, A.~J. Scott, and Carlton~M. Caves.
\newblock Symmetric informationally complete quantum measurements.
\newblock {\em J. Math. Phys.}, 45:2171, 2004.
\newblock http://arxiv.org/abs/quant-ph/0310075v1.

\bibitem{ADF07}
D.M. Appleby, Hoan~Bui Dang, and Christopher~A. Fuchs.
\newblock Physical significance of symmetric informationally-complete sets of
  quantum states, 2007.
\newblock http://arxiv.org/abs/quant-ph/0707.2071v1.

\bibitem{Bus03}
Paul Busch.
\newblock Quantum states and generalized observables: A simple proof of
  gleason's theorem.
\newblock {\em Phys. Rev. Lett.}, 91:120403, 2003.

\bibitem{CFMR04}
Carlton~M. Caves, Christopher~A. Fuchs, Kiran~K Manne, and Joseph~M. Renes.
\newblock Gleason-type derivations of the quantum probability rule for
  generalized measurements.
\newblock {\em Foundations of Physics}, 34:193, 2004.

\bibitem{Spe08}
Robert~W. Spekkens.
\newblock Negativity and contextuality are equivalent notions of
  nonclassicality.
\newblock {\em Phys. Rev. Lett.}, 101:020401, 2008.
\newblock http://arxiv.org/abs/quant-ph/0710.5549v3.

\bibitem{Har04}
Lucien Hardy.
\newblock Quantum ontological excess baggage.
\newblock {\em Stud. Hist. Philos. M. P.}, 35:267, 2004.

\bibitem{BHS93}
Paul Busch, K.-E. Hellwig, and W.~Stuple.
\newblock On classical representations of finite-dimensional quantum systems.
\newblock {\em Internation Journal of Theoretical Physics}, 32:399, 1993.

\bibitem{Cho69}
G.~Choquet.
\newblock {\em Lectures on Analysis}.
\newblock Benjamin, New York, 1969.

\bibitem{ks67}
S.~Kochen and E.~Specker.
\newblock The problem of hidden variables in quantum mechanics.
\newblock {\em J. Math. Mech.}, 17:59, 1967.

\bibitem{Ish95}
Chris~J. Isham.
\newblock {\em Lectures on Quantum Theory: Mathematical and Structural
  Foundations}.
\newblock Imperial College Press, London, 1995.

\bibitem{Spe05}
Robert~W. Spekkens.
\newblock Contextuality for preparations, transformations, and unsharp
  measurements.
\newblock {\em Phys. Rev. A}, 71:052108, 2005.

\bibitem{HR07}
Nicholas Harrigan and Terry Rudolph.
\newblock Ontological models and the interpretation of contextuality, 2007.
\newblock http://arxiv.org/abs/0709.4266.

\bibitem{KL98}
E.~Knill and R.~Laflamme.
\newblock Power of one bit of quantum information.
\newblock {\em Phys. Rev. Lett.}, 81:5672, 1998.

\bibitem{Vid03}
G.~Vidal.
\newblock Efficient classical simulation of slightly entangled quantum
  computations.
\newblock {\em Phys. Rev. Lett.}, 91:147902, 2003.

\bibitem{SC99}
Rudiger Schack and Carlton Caves.
\newblock Classical model for bulk-ensemble nmr quantum computation.
\newblock {\em Phys. Rev. A}, 60:4354, 1999.

\bibitem{KLBR08}
R.~Kaltenbaek, J.~Lavoie, D.~N. Biggerstaff, and K.~J. Resch.
\newblock Quantum-inspired interferometry with chirped laser pulses.
\newblock {\em Nature Physics}, 4:864, 2008.

\end{thebibliography}

\end{document}